\documentclass[twocolumn]{aastex631}

\usepackage{graphics}

\shorttitle{Wideband timing of PPTA UWL data}
\shortauthors{Curyło et al.}

\begin{document}

\title{Wideband timing of the Parkes Pulsar Timing Array UWL data}

\correspondingauthor{Ma\l{}gorzata Cury\l{}o}
\email{mcurylo@astrouw.edu.pl}

\author{Ma\l{}gorzata Cury\l{}o}
\affiliation{Astronomical Observatory, University of Warsaw, Aleje Ujazdowskie 4, 00-478 Warsaw, Poland}

\author{Timothy T. Pennucci}
\affiliation{Institute of Physics, Eötvös Loránd University, Pázmány
P.s. 1/A, 1117 Budapest, Hungary}

\author{Matthew Bailes}
\affiliation{Centre for Astrophysics and Supercomputing, Swinburne University of Technology, P.O. Box 218, Hawthorn, Victoria 3122, Australia}

\author{N. D. Ramesh Bhat}
\affiliation{International Centre for Radio Astronomy Research, Curtin University, Bentley, WA 6102, Australia}

\author{Andrew D. Cameron}
\affiliation{Centre for Astrophysics and Supercomputing, Swinburne University of Technology, P.O. Box 218, Hawthorn, Victoria 3122, Australia}
\affiliation{Australian Research Council Centre of Excellence for Gravitational Wave Discovery (OzGrav), Australia}

\author{Shi Dai}
\affiliation{School of Science, Western Sydney University, Locked Bag 1797, Penrith, NSW 2751, Australia}

\author{George Hobbs}
\affiliation{CSIRO Astronomy and Space Science, Australia Telescope National Facility, PO Box 76, Epping NSW 1710, Australia}

\author{Agastya Kapur}
\affiliation{CSIRO Astronomy and Space Science, Australia Telescope National Facility, PO Box 76, Epping NSW 1710, Australia}
\affiliation{Department of Physics and Astronomy and MQ Research Centre in Astronomy, Astrophysics and Astrophotonics, Macquarie University, NSW 2109, Australia}

\author{Richard N. Manchester}
\affiliation{CSIRO Astronomy and Space Science, Australia Telescope National Facility, PO Box 76, Epping NSW 1710, Australia}

\author{Rami Mandow}
\affiliation{CSIRO Astronomy and Space Science, Australia Telescope National Facility, PO Box 76, Epping NSW 1710, Australia}
\affiliation{Department of Physics and Astronomy and MQ Research Centre in Astronomy, Astrophysics and Astrophotonics, Macquarie University, NSW 2109, Australia}

\author{Matthew T. Miles}
\affiliation{Centre for Astrophysics and Supercomputing, Swinburne University of Technology, P.O. Box 218, Hawthorn, Victoria 3122, Australia}
\affiliation{Australian Research Council Centre of Excellence for Gravitational Wave Discovery (OzGrav), Australia}

\author{Christopher J. Russell}
\affiliation{CSIRO Astronomy and Space Science, Australia Telescope National Facility, PO Box 76, Epping NSW 1710, Australia}

\author{Daniel J. Reardon}
\affiliation{Centre for Astrophysics and Supercomputing, Swinburne University of Technology, P.O. Box 218, Hawthorn, Victoria 3122, Australia}
\affiliation{Australian Research Council Centre of Excellence for Gravitational Wave Discovery (OzGrav), Australia}

\author{Ryan~M.~Shannon}
\affiliation{Centre for Astrophysics and Supercomputing, Swinburne University of Technology, P.O. Box 218, Hawthorn, Victoria 3122, Australia}
\affiliation{Australian Research Council Centre of Excellence for Gravitational Wave Discovery (OzGrav), Australia}

\author{Ren\'ee Spiewak}
\affiliation{Centre for Astrophysics and Supercomputing, Swinburne University of Technology, P.O. Box 218, Hawthorn, Victoria 3122, Australia}
\affiliation{Jodrell Bank Centre for Astrophysics, School of Physics and Astronomy, The University of Manchester, Manchester, M13 9PL, UK}

\author{Andrew Zic}
\affiliation{CSIRO Astronomy and Space Science, Australia Telescope National Facility, PO Box 76, Epping NSW 1710, Australia}
\affiliation{Department of Physics and Astronomy and MQ Research Centre in Astronomy, Astrophysics and Astrophotonics, Macquarie University, NSW 2109, Australia}

\author{Xing-Jiang Zhu}
\affiliation{Advanced Institute of Natural Sciences, Beijing Normal University, Zhuhai 519087, China}

\begin{abstract}

In 2018 an ultra-wide-bandwidth low-frequency (UWL) receiver was installed on the 64-m Parkes Radio Telescope enabling observations with an instantaneous frequency coverage from 704 to 4032\,MHz. Here, we present the analysis of a three-year data set of 35 millisecond pulsars observed with the UWL by the Parkes Pulsar Timing Array (PPTA), using wideband timing methods. The two key differences compared to typical narrow-band methods are, firstly, generation of two-dimensional templates accounting for pulse shape evolution with frequency and, secondly, simultaneous measurements of the pulse time-of-arrival (ToA) and dispersion measure (DM). This is the first time that wideband timing has been applied to a uniform data set collected with a single large-fractional bandwidth receiver, for which such techniques were originally developed. As a result of our study, we present a set of profile evolution models and new timing solutions including initial noise analysis. Precision of our ToA and DM measurements is in the range of 0.005 $-$ 2.08\,$\mu$s and (0.043$-$14.24)$\times10^{-4}$\,cm$^{-3}$\,pc, respectively, with 94\% of the pulsars achieving a median ToA uncertainty of less than 1 $\mu$s. 

\end{abstract}


\section{Introduction} \label{sec:intro}

Pulsar timing array (PTA) experiments provide extraordinary means to study a wide range of physical phenomena across nearly all branches of physics and astronomy. These include characteristics of neutron stars themselves but can also relate to solar system dynamics, general relativity or nHz gravitational waves (GWs) generated by various processes such as supermassive black hole inspirals or cosmic strings (e.g. \citealp{Burke-Spolaor19,Vallisneri20}). However, as all of the above processes may have a very subtle effect on timing measurements (of the order of several ns), an increase of the precision and accuracy is a vital element of current PTA efforts. The three major pillars of PTA, working under a joint venture as the International Pulsar Timing Array: IPTA; \citealp{Manchester13}, are the European Pulsar Timing Array (EPTA; \citealp{Desvignes16}), the North American Nanoherthz Observatory for Gravitational waves (NANOGrav; \citealp{McLaughlin13}) and the Parkes Pulsar Timing Array (PPTA; \citealp{Hobbs2013}). They have also been recently joined by the Indian Pulsar Timing Array (InPTA; \citealp{Ashis19, Tarafdar22}) and are supported by the Chinese Pulsar Timing Array (CPTA; \citealp{Lee16}) and MeerKAT Pulsar Timing Array (MPTA; Miles et al. 2022, submitted). 

There are a number of possible improvements that can be applied to observational strategies, instrumentation and analysis techniques, such as increasing the number of pulsars in the array and the cadence of observations, or enlarging telescope apertures. In particular, several recent projects and facilities such as the Five Hundred Meter Aperture Spherical Telescope (FAST; \citealp{Hobbs19}) and the Canadian Hydrogen Intensity Mapping Experiment (CHIME/Pulsar Project; \citealp{Amiri21,Chime22}) will soon join PTA efforts and significantly increase available observing time and collecting area. 

Another approach is to utilize wideband receivers. In the first instance, the uncertainty of a timing measurements $\sigma$ depends on the observing system and is described by the radiometer equation \citep{Lorimer04}:
\begin{equation}\label{eq:radiometer}
    \sigma \propto \frac{T_{\rm sys}}{A_{\rm eff}}(\tau \Delta f)^{1/2}
\end{equation}
where $T_{\rm sys}$ is the system's temperature, $A_{\rm eff}$ effective aperture, $\tau$ integration time and $\Delta f$ is the bandwidth. Apart from improvements based on Eq. \ref{eq:radiometer}, wideband receivers will also significantly broaden our capabilities of studying processes related to the interstellar medium (ISM), such as scintillation and DM variability.
Currently, there are two telescopes with large instantaneous frequency coverage used for PTA observations: Effelsberg 100-m Telescope with 600$-$3000\,MHz ultra-broadband (UBB) receiver\footnote{{\hyperlink{https://www3.mpifr-bonn.mpg.de/staff/pfreire/BEACON.html}{https://www3.mpifr-bonn.mpg.de/staff/pfreire/BEACON.html}}} and Parkes Radio Telescope (Murriyang) with ultra-wide-bandwidth low-frequency (UWL) receiver covering the largest frequency range of 704$-$4032\,MHz \citep{Hobbs20}. Installation of a new wideband receiver is also planned for MeerKAT \citep{Kramer16} and ultra-wideband (UWB) feed at the Green Bank Telescope used by NANOGrav has recently entered a commissioning phase \citep{Bulatek20}.   

In order to fully benefit from wideband observations it is important to adjust current analysis techniques. Growing apertures, observation time and frequency coverage will significantly increase computational requirements making processing of long PTA data sets arduous. Moreover, as many millisecond pulsars exhibit an intrinsic profile evolution with frequency, pulse shape can change drastically between the extreme ends of the band. These two effects, known collectively as the $large~bandwidth~problem$ are the main factors driving the development of new wideband timing methods. 

In general, the procedure of timing analysis consists of measuring the observed pulse times-of-arrival (ToAs) via template-matching technique which are then compared with the timing model. The model is supposed to accurately predict the periodicity of the pulsar's radio emission, fitting for parameters such as spin period and its derivative, position and proper motion, or binary parameters if applicable. Differences between the observations and the model (the residuals), may manifest as either white or red noise (systematic and time correlated, respectively). In the former case, the main contributors are radiometer noise and jitter \citep{Helfand75,Shannon14}, while for the latter these could be pulsar intrinsic spin noise \citep{Shannon10}, dispersion measure (DM) variations or gravitational waves. In other words, the sensitivity of the PTA depends on the accuracy of both the ToA measurement (and so on the template used) and timing solutions, including a proper characterization of noise sources. 

Typically, template-matching is performed using a frequency-averaged profile template which will deviate from the true pulse shape at different frequencies. In order to mitigate this problem, it is common to divide the band and measure the so called sub-banded ToAs with one average template per sub-band and/or use additional parameters in the timing model (FD parameters). The wideband timing techniques (presented by \citealp{Pennucci14} and \citealp{Liu14}) offer an elegant and more direct solution by generation of a two-dimensional profile template maintaining frequency resolution (model of the profile evolution) and a simultaneous measurement of one ToA and DM at a reference frequency for the whole band. In particular, wide-band timing methods from \cite{Pennucci14} were applied to the 12.5 year data set from NANOGrav \citep[hereafter NG12.5]{Alam21} and directly compared with a corresponding narrow-band release \citep{Alam21b}. It was shown that the number of ToAs was reduced by a factor of 33, while maintaining a similar level of precision in terms of timing model and noise parameters (at least $2\sigma$ agreement). 
NG12.5 have also reported improvement of timing results by 10-15\% for pulsars which were impaired by stronger environmental effects such as high DM and scintillation. Similarly, \cite{Nobleson22} presented a wideband analysis of five pulsars observed with the upgraded Giant Metrewave Radio Telescope (uGMRT) at low frequencies between 300 and 500\,MHz. Apart from showing the aforementioned consistency between results obtained with wide- and narrow-band methods, \cite{Nobleson22} also emphasize an increased precision of low-frequency DM measurements and prove wideband timing methods to be beneficial even for small fractional bandwidths. Other examples of implementations of methods presented by \cite{Pennucci14} can be found in \cite{Fonseca21} and \cite{Sharma22}, where the analyses focus on detailed binary parameter estimation and study of new PTA-candidate pulsars, respectively.

Here, we present wideband timing of 35 millisecond pulsars observed with the UWL by the PPTA using PulsePortraiture \citep{Pennucci16,Pennucci14}. This is the first time such analysis is performed on observations gathered by a single ultra-wide-bandwidth receiver with instantaneous fractional bandwidth of approximately 6:1. It is also the first study of the new PPTA UWL data set in general (note, however, that UWL observations were used to estimate the DM for the timing analysis of the previous data release; \citealp{Reardon21}). Continuous frequency coverage of an ultra-wide band and utilization of PulsePortraiture allowed us to describe the evolution of pulse profiles and subsequently measure ToAs and DMs with raw uncertainties at least two times smaller then those obtained for the previous data set (\citealp{Kerr20}, hereafter PPTA DR2). 

The paper is organized as follows. In Section \ref{sec:obs} we describe the UWL receiver, observation strategies and the new data set. In Section \ref{sec:methods} we present data preparation procedures and the principles of wideband timing and noise analysis. Section \ref{sec:results} contains our results and their discussion, including profile evolution and timing models, noise analysis and notes on a few individual pulsars. Summary and conclusions are presented in Section \ref{subsec:summary}.    

\section{Observations} \label{sec:obs}

Observations analysed in this paper were collected between November 2018 and March 2022 with the 64-m Parkes Radio Telescope (Murriyang) located in New South Wales, Australia. The starting date marks the transition from the previous observing systems to the UWL which has been continuously carrying out all observations ever since. Full description and technical details of the new receiver and associated systems can be found in \cite{Hobbs20} and below we introduce it only briefly.

Both the feed and low-noise amplifiers are cryogenically cooled and maintain low, 22\,K temperature for the majority of the band. Pre-processed data consists of 26 critically sampled sub-bands, each 128 MHz wide, continuously covering 704$-$4032 MHz frequency range. For timing purposes, we use data coherently de-dispersed and folded into 1024 phase bins by the DSPSR suite \citep{Straten11} on the Medusa GPU cluster. Noteworthy, Medusa is the only processing system presently used as opposed to seven employed in the previous data release PPTA DR2.

For flux calibration purposes we observe two bright sources (PKS\,B1934$-$638 and PKS\,B0407$-$658) approximately once per session. Additionally, each pulsar observation is preceded (and sometimes also followed) by a two-minute injection of a noise diode signal. 

The list of all 35 pulsars observed with the UWL for PPTA project (P456) is presented in Table \ref{table:results}. Observations are carried out with a standard cadence of approximately 1$-$3 weeks. 25 of the sources were included in the previous data release, while the remaining 10 were added to the array between 2018 and 2020 and are currently being reviewed as potential candidate pulsars for PTA.

\section{Methods} \label{sec:methods}

\subsection{Wideband timing} \label{subsec:wideband}
We refer the reader to \cite{Pennucci14, Pennucci19} for a detailed explanation of the wideband timing procedures, however, a brief summary is presented below. 

We first make a smooth, noise-free, average (in time and frequency) profile from one, highest signal to noise ratio (S/N) observation, which is then used iteratively to align several tens of epochs comprising our initial frequency resolved template. This step is similar for both narrow- and wide-band techniques, however, the standard phase shift between the profiles and the noise-free template is additionally a function of frequency, i.e. is described by a dispersion law:
\begin{equation}
    \label{eq:phase}
    \phi_{\rm n}(\nu_{\rm n}) = \phi_0 + \frac{\rm K \cdot DM }{\rm P_{\rm S}}(\nu_{\rm n}^{-2} - \nu_{\phi_0}^{-2})
\end{equation}
where $\phi_0$ is the phase offset at reference frequency\footnote{The reference frequency is chosen such that there is zero covariance between DM and $\phi_0$ (see Appendix in \cite{Pennucci14}} $\nu_{\phi_0}$), $\rm{K = 4149.37759336~MHz^2~cm^3~pc^{-1}~s}$ is dispersion constant and $P_{\rm S}$ is the spin period of the pulsar. A collection of these aligned profiles (the portrait) is then decomposed into eigenvectors via principal component analysis (PCA) and the ones with the highest S/N  (along with the mean profile) are used to model the frequency evolution of the profile. At the final step, a spline function is fitted to the projection of each mean substracted portrait profile onto the significant eigenvectors. In short, we can reconstruct the profile shape at any desired frequency by summing up the product of $n_{\rm eig}$ spline functions $S_{\rm i}(\nu)$ and significant eigenvectors $\hat{e_{\rm i}}$ and adding it to the mean profile $\tilde{p}$: 

\begin{equation}
    \mathrm{T(\nu) = \sum_{i=1}^{n_{eig}} S_{i}(\nu)\hat{e_{i}} + \tilde{p}}
\end{equation}

These procedures then allow for a simultaneous measurement of one ToA and DM per whole band (at a reference frequency $\nu_{\phi_0}$) for each observation by minimizing the statistic:

\begin{equation}
    \label{eq:chi2}
    \chi^2 = \sum_{\rm n,k} \frac{|d_{\rm nk} - a_{\rm n}t_{\rm nk}e^{\rm -2\pi ik \phi_{\rm n}}|^2}{\sigma_{\rm n}^2}
\end{equation}
\\
where indices $k$ and $n$ run over Fourier frequencies and channels, respectively, $d_{\rm nk}$ is a one-dimensional (along phase axis) discrete Fourier transform (DFT) of the profile and $t_{\rm nk}$ is the DFT of our template, $a_{\rm n}$ is the amplitude scaling factor and $\sigma_{\rm n}^2$ is the Fourier-domain noise level.

\subsection{Data preparation}\label{subsec:data_prep}

In order to prepare data for our wideband analysis, we first cleaned it from radio-frequency-interference (RFI), calibrated and frequency averaged down to 416 channels (each 8\,MHz wide) by an automatic pipeline (also used in \citealp{Kerr20}). Afterwards, we prepared our portraits by aligning and averaging nearly all available UWL observations as explained in Sec.\,\ref{subsec:wideband}. The two exceptions here were the brightest or faintest sources. In the former case, 10$-$30 observations were sufficient enough to produce high S/N portraits and at the same time allowed for more accurate RFI excision. In case of the faintest sources in the data set, some of the observations were heavily corrupted by RFI or instrumental errors, and due to a much smaller number of observing epochs relative to brighter pulsars they significantly affected the portrait and so were removed. 
We also note, that each portrait was manually checked for RFI and channels with $\rm S/N\sim0$ before modeling procedures because: i) summing up the profiles may bring up previously missed contaminated channels, and ii) any spurious signal or impaired channels can significantly affect the model of profile evolution. Finally, once the templates were obtained, we derived spline models and subsequently measured one ToA and one DM per band for each observation as explained in Sec \ref{subsec:wideband}. All of the above procedures were performed using an open source code\footnote{https://github.com/pennucci/PulsePortraiture/} - PulsePortraiture \citep{Pennucci14}.

The last step before proceeding to timing analysis was to filter out all bad epochs in our data sets. For that purpose, we have defined four conditions which had to be met in order to include particular observation in the further analysis. These were: 
\begin{enumerate}
    \item Observation time: $t_{obs} > 300$\,s.\\
    Nominal length of each observation is 1.1\,hour. There is however, a subset of shorter observations due to technical limitations, e.g. changing weather conditions, finite length of observing session, RFI, etc. We do include those partial observations in our analysis,  only if their length exceeds 300\,s. Such a low threshold allows to include more observations of the brightest pulsars, whereas short ones usually didn't fulfill also the remaining conditions.
    
    \item ToA S/N: $S>25$.\\
    The quality of low S/N observations results in poor estimation of measured ToAs and DMs (high uncertainties). Note however, that the threshold applies to S/N of the wideband ToAs not observations, and is defined as: 
    $S \equiv \sqrt{\sum_{\rm n} S_{\rm n}^2}$, where $S_{n}$\,$\equiv a_{\rm n}\sqrt{\sum_{\rm k} |t_{\rm nk}|^2} / \sigma_{\rm n}$ (for a more detailed description see Appendix A in \citealp{Alam21}).
    
    \item Goodness-of-fit: $\chi^2_{\rm reduced} < 1.25$.\\
    The $\chi^2$ statistics is calculated for each observation, taking into account the model of profile evolution. Large values of this parameter can imply non-curated RFI or low quality profiles. 
    
    \item Highest/lowest frequency ratio: $f_{\rm ratio} > 1.1$.\\
    Observations filtered out based on this condition are either heavily contaminated by RFI or affected by serious instrumental issues, where the signal throughout majority of the band is lost. All observations which we included in our analysis had $f_{\rm ratio} > 2.9$ which corresponds to the effective bandwidth of approximately 2600 MHz (most of observations with $f_{\rm ratio} \sim 2.9$ were cut below 1400 MHz due to strong RFI). For full bandwidth $f_{\rm ratio} = 5.7$. 

\end{enumerate}

This procedure allowed us to remove most of the bad epochs automatically and only a small fraction of individual observations were later flagged manually due to large residuals or DM/ToA uncertainties, which, as expected, occured mainly for low quality profiles. Each epoch which was commented out was additionally checked by eye to determine whether it could be curated and added back to the data set. This procedure was possible only because of a still relatively small number of observations, however we acknowledge the need for more accurate RFI zapping algorithms for future analyses. In general, we used between 80\% to 100\% of available UWL observations for each pulsar. 

Finally, we note that all of the pulsars in the data set were analysed using total intensity profiles (Stokes~I), apart from J0437$-$4715 for which we used the polarimetric invariant profile, as was also done in previous analyses \citep{Straten01, Kerr20}. The invariant interval can be used to avoid additional red noise from polarization calibration errors and to reduce dependence of the observation on the parallactic angle \citep{Hotan06}. It is given by \citep{Hotan06}:\\
\begin{equation}
    \rm S_{\rm inv} = I^2 - Q^2 - U^2 - V^2
\end{equation}\\
where $I, Q, U, V$ are Stokes parameters. It can be used in the case of the least polarized sources or when large part of the emission is unpolarized, which is why we applied it only to J0437-4715.
Our final data set consists of frequency dependent profile templates (portraits), spline models of profile evolution and a list of measured ToAs and DMs with their uncertainties for each pulsar. 

\subsection{Timing solutions and noise analysis}
As explained in Section \ref{subsec:wideband}, each wideband measurement consists of one ToA and a corresponding DM at the time of the observation along with their uncertainties. It is possible to obtain a wideband ToA without fitting for the DM, however that would still necessitate providing an external and precise measurement of the DM at the time of the observation in order to properly align the data with the template. In either case, further analysis of the timing and noise models also requires using both measurements per observation, as opposed to sub-banded ToAs, single wideband ToA does not contain the full information of the dispersive delay within the observing band.
In consideration of that, in our work we used two packages, i.e. Tempo \citep{Nice15} for pulsar timing and ENTERPRISE \citep{Ellis19} for noise modeling, where the new wideband likelihood has been already been implemented (alternatively, wideband analysis can also be performed in PINT; \citealp{PINT}). The mathematical description of these implementations can be found in Appendix B in \citealp{Alam21}. Below, we present all details of obtaining our wideband timing solutions and corresponding noise models along with a general overview of the augmented likelihood. 

We started with initial timing solutions obtained for the PPTA DR2 \citep{Goncharov21} applying the following changes. First, we removed all noise parameters, as they have to be derived again using the new wideband likelihood. We have also updated the clock standard and solar system ephemeris to TT(BIPM2019) and DE438, respectively. Another change was applied to the choice of binary models. The majority of pulsars with binary companions from PPTA DR2 were fitted with the T2 model which is available only in the Tempo2 package \citep{Hobbs06}. Here instead, for most pulsars we used the ELL1 model (adequate for low eccentricity orbits; \citealp{Lange01}), for one pulsar (J1643-1224) we used DD model (to  include measurement of eccentricity; \citealp{Damour92}), for two pulsars (J0437$-$4715, J1713$+$0747) we used the DDK model (allowing for measurements of annual-orbital parallax; \citealp{Kopeikin95,Kopeikin96}), and four pulsars (J1017$-$7156, J1022$+$1001, J1545$-$4550, J1600$-$3053) were fitted with ELL1H model (which includes modelling of Shapiro delay; \citealp{Freire10}).  

The next element of our timing analysis is modeling of the DM variability.  We used the DMX model which assumes that the DM is constant within chosen time intervals (e.g. a fraction of a day or several days), and models the subsequent changes in a piecewise-constant manner. The choice of the DMX epoch length depends on various factors, e.g. observing strategy, desired precision or expected ISM/solar wind variability. In this work we have applied DMX epochs between 1 to 60 days, depending on the number of measurements in order to avoid overfitting. 
Each DM from our wideband measurements was then used as a prior on the DMX value in the corresponding epoch. In the case there were more ToAs for a given epoch, the prior was calculated from a weighted average. 

Finally, noise present in the timing residuals was modelled using Bayesian inference implemented within ENTERPRISE. 

To account for the white noise, ToA uncertainties for each measurement $\sigma_{\rm j}^{\rm ToA}$ were modified by 2 Gaussian-noise components:
\begin{equation}
    \rm \sigma_{\rm j}^2 = (EFAC~\sigma_{\rm j}^{ToA})^2 + EQUAD^2
\end{equation}
where EFAC encapsulates unknown systematic errors associated with observing systems and analysis, and EQUAD added in quadrature characterizes any additional, system-independent white noise.

Within the wideband likelihood, two new white noise components have been introduced: DMEFAC and DMEQUAD which are analogous to these described above but are applied to DM uncertainties. In addition to that, we also used a new DMJUMP parameter, which can be referred to the standard JUMP parameter (used to account for phase offsets of unmodeled profile evolution), but here it represents the offset between the mean wideband DM and individual wideband DM measurements. In other words, it accounts for the ambiguity of determining the absolute DM. Finally, a standard ECORR parameter can be omitted in the wideband analysis, as it is describing the correlation between sub-banded ToAs naturally not present here (noise sources contributing to ECORR are absorbed by EQUAD in the wideband likelihood).

We set narrow Gaussian EFAC and DMEFAC priors on 1.00\,$\pm$\,0.25. Priors for the other parameters are drawn from uniform distributions given by: \\
$\mathrm{log_{10}{(EQUAD)} \in [-8.5,-5.0]}$,\\
$\mathrm{log_{10}{(DMEQUAD)} \in [-7.0,0.0}]$,\\
$\mathrm{log_{10}{(DMJUMP)} \in [-0.01,0.01]}$.

\begin{deluxetable*}{c c c c c c c c c c c c}[ht!]
\tablecaption{Summary of timing analysis for 35 pulsars. 25 of the listed pulsars are high priority sources observed as part of the PPTA project, while the remaining 10 (marked with a star) have been added to the array after the installation of UWL, between 2018 and 2020. The 10th column shows the S/N of the pulse portraits.}
\label{table:results}
\tablehead{
\colhead{Pulsar} & 
\colhead{Span} & 
\colhead{DM} & 
\colhead{P} & 
\colhead{ToAs} & 
\colhead{Pars} &
\colhead{RMS} &
\colhead{Med $\sigma_{\rm ToA}$}& 
\colhead{Med $\sigma_{\rm DM}$}& 
\colhead{S/N} & 
\colhead{$\mathrm{n_{eig}}$} & 
\colhead{Figure} \\
\colhead{} & 
\colhead{[yr]} & 
\colhead{[\,cm$^{-3}$\,pc]} & 
\colhead{[ms]} & 
\colhead{} & 
\colhead{} &
\colhead{$\rm[\mu s]$} &
\colhead{$\rm[\mu s]$} & 
\colhead{[$\times10^{-4}$\,cm$^{-3}$\,pc]} & 
\colhead{} & 
\colhead{} & 
\colhead{}
}
\startdata
J0030$+$0451*	 & 3.1 & 4.33 & 4.87 & 34   & 5 & 0.784 & 0.732   & 4.423	& 427	& 1    &\ref{fig:plot0}	\\
J0125$-$2327*	 & 3.1 & 9.60 & 3.68 & 105  & 13  &  0.529  & 0.129	 & 0.974 & 5503 & 3   &\ref{fig:plot0}	\\
J0348$+$0432*	 & 2.7 & 40.47 & 39.12 & 22   & 10 & 2.511  & 2.080	 & 14.24 & 211	& 0    & \ref{fig:plot0}	\\
J0437$-$4715	 & 3.3  & 2.64 & 5.76 & 222  & 13 & 0.195 & 0.005 & 0.043  & 18191	& 6 & \ref{fig:plot0}	\\ 
J0613$-$0200	 & 3.4 & 38.78 & 3.06 & 76  & 13  & 0.269 & 0.158 & 0.807 & 1633 & 2 &\ref{fig:plot0}	\\
J0614$-$3329*	 & 3.1 & 37.05 & 3.15 & 90  & 10	& 0.813	 & 0.741   & 4.171	& 490	& 1    &\ref{fig:plot0}	\\ 
J0711$-$6830	 & 3.3 & 18.41 & 5.49 & 157  & 7  & 0.435 & 0.550	 & 3.211 & 3927 & 2 &\ref{fig:plot1}	\\ 
J0900$-$3144*  & 2.7 & 75.61 & 11.11 & 65  & 12	& 0.931	 & 0.612	 & 3.421 	& 2376	& 2    &\ref{fig:plot1}	\\
J1017$-$7156	 & 3.3 & 94.22 & 2.34 & 173  & 16 & 0.133  & 0.120   & 0.684	& 2090 & 2    &\ref{fig:plot1}	\\ 
J1022$+$1001	 & 3.4 & 10.25 & 16.45 & 59  & 12 & 0.566	& 0.254   & 1.517	& 6189	& 3   &\ref{fig:plot1}	\\ 
J1024$-$0719	 & 3.3 & 6.48 & 5.16 & 50 & 9 & 0.771 & 0.619	 &  4.105	& 1516	& 2    &\ref{fig:plot1}	\\ 
J1045$-$4509	 & 3.3 & 58.15 & 7.47 & 54 & 12  &  1.316  & 0.781	 & 4.173 & 2408	& 2 &   \ref{fig:plot1}	\\ 
J1125$-$6014	 & 3.3 & 52.93 & 2.63 & 97 & 15	& 0.203	 & 0.119	 & 0.744	& 970	& 2     &\ref{fig:plot2}	\\ 
J1446$-$4701	 & 3.3 & 55.83 & 2.19 & 73 & 13 & 0.665	& 0.425	 & 2.741	& 431	& 0    &\ref{fig:plot2}	\\ 
J1545$-$4550	 & 3.3 & 68.39 & 3.58 & 100  & 15  & 0.327 & 0.212  & 2.144  & 876	& 2    &\ref{fig:plot2}	\\ 
J1600$-$3053	 & 3.3 & 52.33 & 3.60 & 53  & 16 & 0.263  & 0.113	 & 0.832	& 2694	& 2    &\ref{fig:plot2}	\\ 
J1603$-$7202	 & 3.3 & 38.04 & 14.84 & 105  & 12 & 0.706 & 0.361 & 2.129 & 6643 & 2 &  \ref{fig:plot2} \\ 
J1643$-$1224	 & 3.3 & 62.41 & 4.62 & 42  & 14	 & 1.25 & 0.286 & 1.459  & 2870	& 2 &   \ref{fig:plot2}	\\ 
J1713$+$0747	 & 2.3 & 15.92 & 4.57 & 43  & 13  & 0.241  & 0.048	 & 0.384 & 9893 	& 2    & \ref{fig:plot3}	\\ 
J1730$-$2304	 & 3.3 & 9.62 & 8.12 & 39  & 6  & 1.244  & 0.287	 & 1.677  & 3660	& 2    &\ref{fig:plot3} \\ 
J1741$+$1351*  & 2.6 & 24.20 & 3.74 & 16 & 10 & 0.460  & 0.345 & 2.640 & 145	& 0 &   \ref{fig:plot3}	\\
J1744$-$1134	 & 3.3 & 3.14 & 4.07 & 88  & 8  & 0.327  & 0.080  & 0.524  & 8670 & 2 &  \ref{fig:plot3}	\\
J1824$-$2452A  & 2.8 & 119.90 & 3.05 & 11   & 5   & 4.481   &  0.123  & 0.710 & 388 & 1 &\ref{fig:plot3} \\
J1832$-$0836	 & 2.7 & 28.20 & 2.72 & 21 & 8  & 0.246 & 0.227	 & 2.107 & 290	& 1    & \ref{fig:plot3}	\\
J1857$+$0943	 & 3.1 & 13.30 & 5.36 & 25  & 10  & 0.391  & 0.257	 & 2.189 	& 1366 & 2    & \ref{fig:plot4}	\\ 
J1902$-$5105*  & 2.7 & 36.25 & 17.40 & 31  & 7  & 2.884  & 0.329 	 & 1.910	& 121 & 0     &\ref{fig:plot4} \\
J1909$-$3744	 & 3.4 & 10.39 & 2.95 & 220  & 15  & 0.231  & 0.027   & 0.176	& 15208	& 1    &\ref{fig:plot4}	\\ 
J1933$-$6211*  & 3.1 & 11.50 & 3.54 & 84  & 12 & 0.752  & 0.764   & 4.842	& 1381	& 1     &\ref{fig:plot4}	\\
J1939$+$2134	 & 2.7 & 71.01 & 1.56 & 12  & 5	& 0.856 & 0.010 & 0.050 	&  2049 & 2 & \ref{fig:plot4}		\\ 
J2051$-$0827*  & 2.6 & 20.73 & 4.51 & 34 & 8	 & 8.643 & 0.680	 & 5.657	& 1462	& 2    &\ref{fig:plot4}	\\
J2124$-$3358	 & 3.3 & 4.59 & 4.93 & 67  & 8  &  1.698  & 0.985	 & 5.760	& 410	& 1   &\ref{fig:plot5}	\\ 
J2129$-$5721	 & 3.3 & 31.85 & 3.73 & 109  & 12	& 0.679   & 0.509 & 3.512  & 1895 	& 1     &\ref{fig:plot5}	\\ 
J2145$-$0750	 & 3.3 & 9.00 & 16.05 & 51 & 13	&  0.823 & 0.186   & 1.179	& 7117	& 2    &\ref{fig:plot5}\	\\ 
J2150$-$0326*  & 2.6 & 20.67 & 3.50 & 26 & 10  & 2.011 & 1.115 & 6.827 & 128  &  0   & \ref{fig:plot5}	\\
J2241$-$5236	 & 3.3 & 11.41 & 2.19 & 151 & 13 & 0.283 & 0.070	 & 0.465 & 8651 	& 2    &\ref{fig:plot5} \\
\enddata
\end{deluxetable*}

\section{Results and discussion} \label{sec:results}

Table \ref{table:results} contains basic pulsar parameters and summarizes the results of our ToA/DM measurements and timing analysis. Apart from nominal DM, orbital period, median ToA/DM uncertainties and RMS of the timing residuals, we also listed the number of ToAs and fitted timing model parameters, S/N of the average portraits, and the number of eigenprofiles as a general characteristics of the profile evolution models. In Appendix\,\ref{sec:AppA} we show all the residual and DM variability plots (referenced in the last column of Table  \ref{table:results}) and in Appendix\,\ref{sec:AppB} we present tables of our timing model parameters.

\subsection{Average portraits}\label{subsec:average_portraits}
\begin{figure}[ht!]
    \includegraphics[width=\columnwidth]{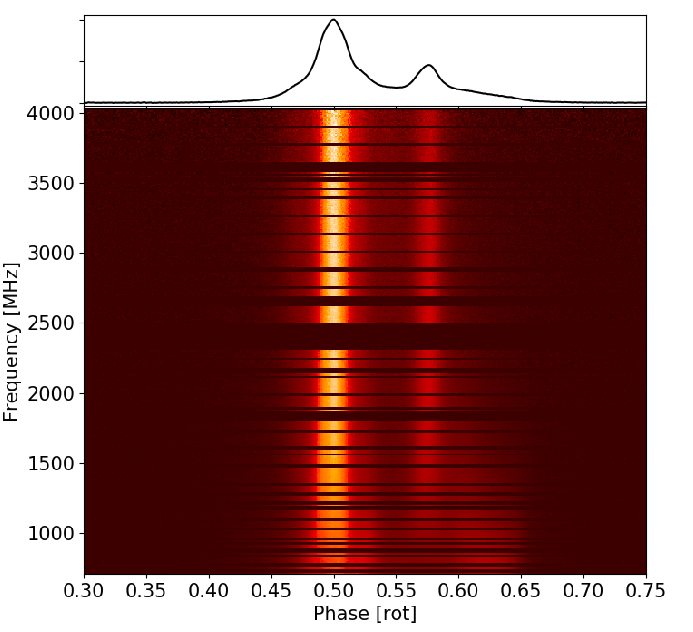}
    \caption{Data portrait of J0125$-$2327 with the mean profile shown in the upper panel.}
    \label{fig:J0125_dp}
\end{figure}

As described in Sec.\,\ref{subsec:wideband}, an average portrait is composed of several tens of aligned and averaged observations. As an example, we show the portrait of J0125$-$2327 in Fig.\,\ref{fig:J0125_dp}.

Each gap in frequency coverage corresponds to zero-weighted channels due to strong RFI. The overall contamination of the whole band is relatively low and the two most significant sources of spurious signal are mobile ($< 1000$ MHz) and WiFi/Bluetooth ($\sim 2400$ MHz) transmissions.

Any residual intensity variations preserved despite time averaging of the portrait are eliminated by the normalization of each frequency channel with respect to the mean profile, and so the amplitude in Fig.\,\ref{fig:J0125_dp} is in arbitrary units. This procedure is meant to ensure that the model correctly describes the intrinsic profile shape changes irrespective to the effects of spectral index or ISM variability (specifically the diffractive scintillation). 

\subsection{Profile evolution models}\label{subsec:profiles}
Profile evolution with frequency is encoded in the spline curve model, and the number of significant eigenvectors (eigenprofiles) can be treated as a proxy for the level of profile evolution complexity. Noteworthy, the later depends not only on the intrinsic profile shape changes, but also on any distortions of the signal (e.g. due to ISM or RFI) and observing system/analysis failures.

When there are no eigenprofiles detected, all measurements are referred to the mean profile. In such a case, there is either none/little profile evolution or S/N for a given pulsar is too low. The first eigenprofile corresponds to a gradient of the mean profile changes (provided it is the only eigenprofile detected), while the second and third may encompass more extreme profile evolution, possible ISM effects such as unmodeled scattering, or issues related to data reduction and analysis. These could be, e.g. misalignment of the profiles comprising the average portrait and inaccuracies in polarization calibration. More than three eigenprofiles can be detected for very high S/N pulsars, but usually they arise from systematic errors or unmitigated RFI. 

The number of eigenprofiles for each pulsar in our data set is listed in Table  \ref{table:results}. The majority of pulsars required two eigenprofiles (19 out of 35), one was detected for eight pulsars and none in case of five sources. The remaining three pulsars were described by three (J0125$-$2327, J1022$+$1001) and six eigenprofiles (J0437$-$4715). There is an evident relation between the complexity of the spline model and S/N of the average portrait, indicating that the quality of observations is one of the leading factors affecting the precision of pulse shape modeling. We discuss this further in Sec.\,\ref{subsec:subbanded}.

\begin{figure}[ht!]
    \includegraphics[width=\columnwidth]{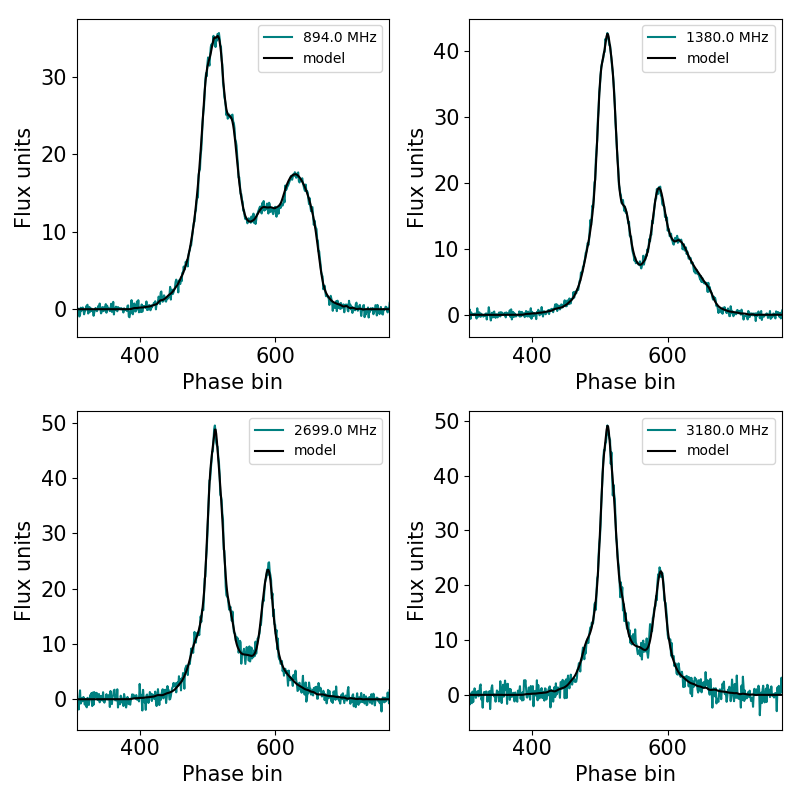}
    \caption{Observed (teal) and reconstructed by the spline model (black) pulse profiles at four frequencies throughout the UWL band for J0125$-$2327 (upper left to lower right: 894, 1380, 2699 and 3180 MHz).}
    \label{fig:J0125_model_obs}
\end{figure}

Below, as an example, we present the profile evolution model for J0125$-$2327. This is a new pulsar to the PPTA discovered in the Green Bank Northern Celestial Cap Pulsar Survey \citep{Stovall14}, characterized by a profile with two leading components exhibiting strong evolution with frequency. Fig.\,\ref{fig:J0125_model_obs} shows the observed profile shapes at four frequencies throughout the UWL band extracted from the average portrait (teal) and modeled pulse profiles (black). Despite obvious complexity of the pulse shape and its evolution, the model is closely tracing all visible details with high accuracy.   

\begin{figure}[ht!]
    \includegraphics[width=\columnwidth]{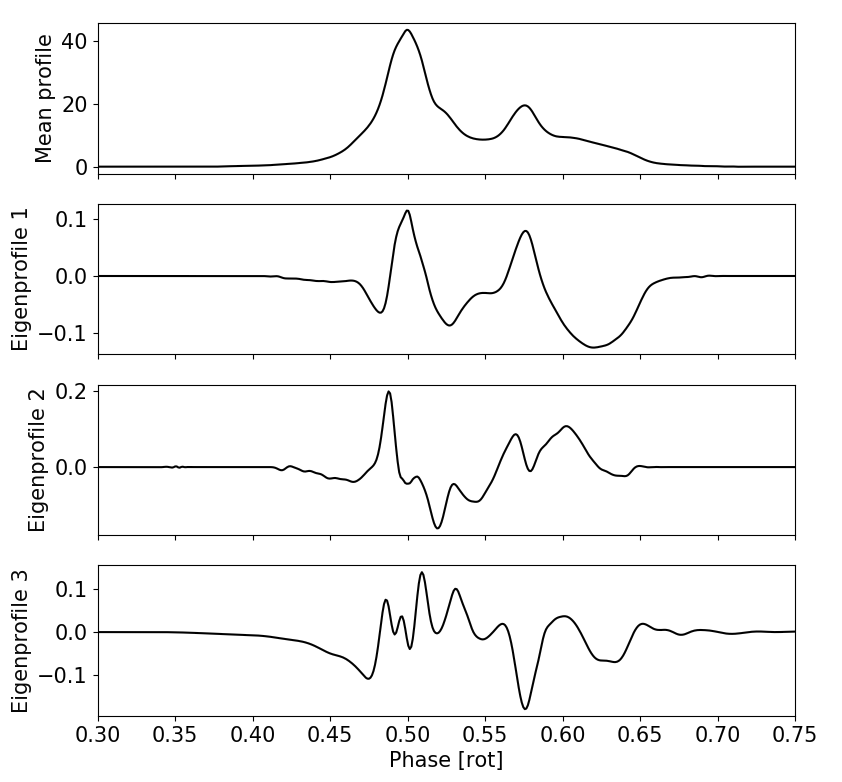}
    \caption{Mean profile and three eigenprofiles for J0125$-$2327. Units at the y-axis are arbitrary.}
    \label{fig:J0125_ep}
\end{figure}

\begin{figure}[ht!]
    \includegraphics[width=\columnwidth]{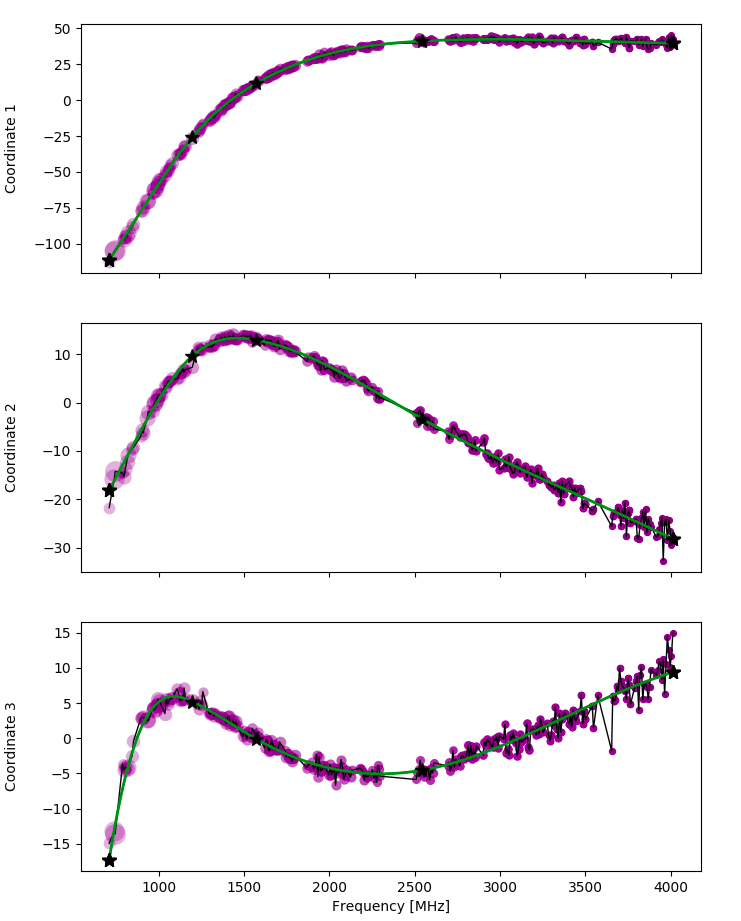}
    \caption{Spline model of profile shape evolution with frequency for J012$5-$2327. Purple points are mean substracted profiles comprising the average portrait projected onto the eigenprofiles. Color scale and size of the points correspond to frequency and S/N of the profile, respectively. The green line represents the spline model of profile evolution.}
    \label{fig:J0125_spline}
\end{figure}

J0125$-$2327 is one of only three pulsars for which we detected three significant eigenprofiles (shown in Fig.\,\ref{fig:J0125_ep}). Each eigenprofile is smooth and well resolved with high S/N. Additionally, coordinate curves comprising the spline model shown in Fig.\,\ref{fig:J0125_spline} are clear and do not exhibit any signs of RFI contamination. In fact, the presented model can be considered as a flagship example given the complexity of the pulse shape and the level of detail captured.

In the case of the majority of our pulsars which are described by two, high S/N eigenprofiles, their spline models do not indicate any substantial issues and fit observed profiles at various frequencies with a comparable level of accuracy as shown above. Five out of 35 pulsars with no significant eigenvectors have portrait $\rm S/N < 430$, usually with low flux measured above 2500 MHz. Similarly, subtle profile shape characteristics and its changes can be lost for models with on only one eigenprofile, especially in the upper part of the band where the flux of most of pulsars is usually lower. Two extreme examples of such sources are J0030$+$0451 and J2129$-$5721, both with large spectral indices $\alpha_{\rm J0030} = - 2.4$ and $\alpha_{\rm J2129} = - 3.9$ \citep{Spiewak22}. In order to capture any profile evolution, and more accurately predict the pulse shapes at lower frequencies, we decided to remove the upper part of the band above 3000-3500 MHz. This resulted in a detection of a previously not present eigenprofile for J2129$-$5721 and a much better fit to the observed profile shapes in case of both pulsars. 

Finally, we would like to stress, that the fact that all pulsars we studied required only zero to three eigenprofiles indicates excellent quality of UWL observations maintained throughout the whole band (free of substantial instrumental, calibration or profile alignment errors). In addition to that, our study demonstrates a notable efficiency of PulsePortraiture, where even complex pulse shape changes can be captured and described by a reasonably simple model.

\subsection{Sub-banded ToAs}\label{subsec:subbanded}
\begin{figure*}[ht!]
    \includegraphics[width=\hsize]{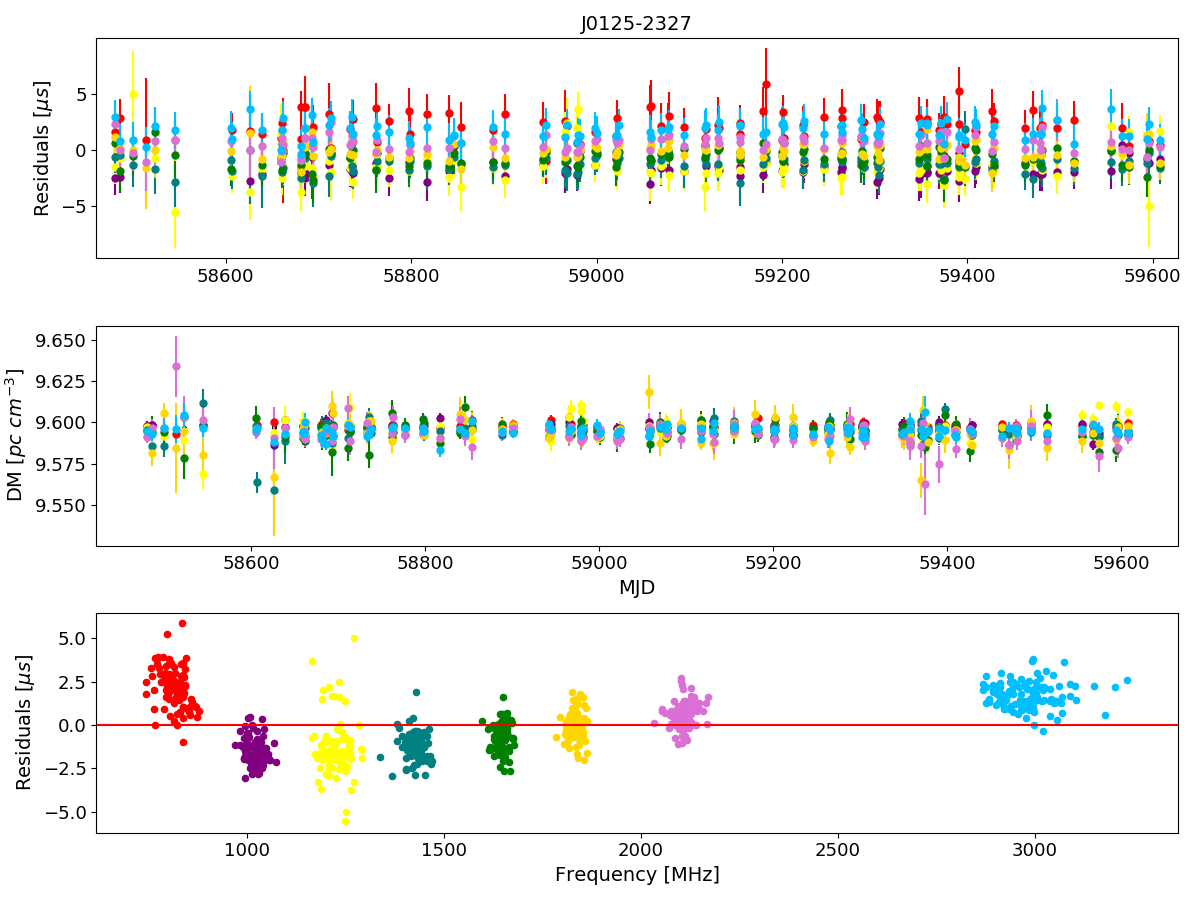}
    \caption{Sub-banded residuals and DM measurements for J0125$-$2327. The bottom panel shows residuals as a function of frequency. Each color corresponds to one of the frequency bands.}
    \label{fig:J0125_sb}
\end{figure*}

Although wideband timing methods offer a number of improvements such as simplification of timing models and lowering data volumes, a parallel analysis of sub-banded ToAs may help reveal frequency-dependent effects related either to physical phenomena, such as DM variations, or technical issues. At the same time, wideband profile evolution modeling is also sensitive to instrumental problems (for instance, resulting in spurious eigenprofiles) and was shown to improve timing of scintillating and weak pulsars \citep{Alam21}. Moreover, comparison of noise models from narrow- and wideband data sets can help discriminate various sources of noise, because of a different set of parameters. Therefore, synergy between the two approaches would be highly beneficial for high precision pulsar timing, allowing for better characterizations of individual pulsars, ISM and their models. 

Following the above arguments, we divided all UWL observations into eight sub-bands and calculated wideband ToAs separately in each band, using the model derived for the whole band. The first six bands have an equal width of 208 MHz and cover frequency range from 704 MHz to 1952 MHz, while bands seven and eight are broader (416 MHz and 1664 MHz, respectively) providing a more even distribution of S/N. 

In Fig.\,\ref{fig:J0125_sb} we present sub-banded residuals as a function of time and frequency (top and bottom panels, respectively) and DM measurements for pulsar J0125$-$2327 as an example. This is one of the best timed sources in our sample, with $\sigma_{\rm ToA} = 129$\,ns, a portrait with high $\rm S/N = 5503$ and a detailed model containing three, well defined eigenprofiles presented in Sec.\,\ref{subsec:profiles}. Nevertheless, the bottom panel of Fig.\,\ref{fig:J0125_sb} shows an evident frequency drift of the residuals with an amplitude of approximately $3.79~\mu$s. After a thorough inspection we found such drift in all of our pulsars, however with varying significance. Median amplitude of the drift is about $1.65~\mu$s and in most cases it lies below the phase bin resolution. 

This leads to the first hypothesis: if pulse shape changes occur at such small scales, then the spline models may not be able to capture their frequency evolution in full detail resulting in frequency dependence of the residuals. However, we find that scenario unlikely, as power spectra of the average portraits show an expected exponential drop of the signal with harmonic number, while any unresolved features would remain above the white noise at all harmonics. 

More importantly, phase resolution can be a limiting factor in the process of profile alignment for the average portraits. As noted in \citealp{Pennucci19}, intrinsic profile evolution is entangled with the absolute DM or its variations and the alignment of the portraits may be accurate up to one phase bin level. Furthermore, due to the dispersive law, any inaccuracies in the DM measurements will induce profile smearing on timescales increasing with bandwidth, thus the effect will be more prominent for ultra-wide-band receivers, such as the UWL. The rotational period of J0125-2327 is 3.68\,ms, which means that the amplitude of the frequency drift seen in Fig. \ref{fig:J0125_sb} is equal to approximately one phase bin, further supporting this scenario. The corresponding $\delta$DM that would induce this level of drift across the UWL band is $\rm \sim5\times10^{-4}\,cm^{-3}$\,pc, which is comparable with the median DM uncertainty of $\rm \sim1\times10^{-4}\,cm^{-3}$\,pc.

Another possibility is that these drifts are induced or enhanced by timing and noise modeling systematics or during the decomposision of the average portraits with PCA, which can be supported by the form of their frequency dependence. Of course, all of the above hypotheses do not have to be mutually exclusive and the observed frequency drift may very well be a function of multiple factors. 

Four pulsars, where the drift is larger than the bin resolution have either zero or one eigenprofile (J1446$-$4701, J2124$-$3358 and J2129$-$5721, respectively), have a broad profile possibly affected by scattering (J1045$-$4509) and a weak upper-most part of the band which was cut out during modeling (J2129$-$5721). In these cases we expect some of the information on the profile evolution to be missing or obscured, and therefore a few possible improvements would require observations with higher S/N and detailed modeling of the ISM effects. 

This work is the first to report and delineate the unmodeled frequency dependence in the timing residuals from wideband timing, although PulsePortraiture and wideband techniques have been already used for multiple and diverse studies as was mentioned in Sec.\,\ref{sec:intro}. The phenomenon is naturally present in narrow-band analyses which use frequency-averaged templates and is usually corrected by applying additional FD parameters to the timing model in order to account for profile evolution with frequency. The main goal of the wideband technique is to capture and model these frequency-dependent effects and thus detection of the drift in our wideband data is unsettling. We note however, that a similar trend has been recently detected also in other data sets including wideband analysis from NANOGrav (via private communication). Additionally, timing models from the latest MPTA release included FD parameters despite using frequency evolving templates (Miles et al. 2022, submitted). 

The fact that the drift was observed only now may have multiple reasons. Frequency dependent phenomena will be less pronounced or even absolutely undetectable for narrow-bandwidth observations, and wideband timing methods have, in fact, never been used on data with more than 2 GHz instantaneous frequency coverage and to such extend (nearly all new timing analyses now include wideband methodology). For instance, wideband analysis presented in NG12.5 did not report the need for FD parameters, because the ToAs were effectively calculated only for two sub-bands (as opposed to eight sub-bands shown in Fig. \ref{fig:J0125_sb}), which obscured any such detectable drifts. Similarly, \cite{Tarafdar22} also presented both narrow- and wideband timing of 14 pulsars as the first data release from the InPTA and explicitly showed no need for additional corrections for frequency-dependent effects. However, InPTA analysis was also perfomed on only two sub-bands and produced narrow-band, frequency-resolved templates by iteratively fitting for DM to the multi-band observations. Noteworthy, \cite{Tarafdar22} pointed out that differences in denoising the templates lead to a discrepancy in DM estimates obtained by narrow- and wideband techniques, which might be worth further investigation.

Interestingly, parallel narrow- and wideband analyses presented in e.g. NG12.5 and \cite{Nobleson22} yielded remarkable agreement between the results from the two methods. This indicates that frequency-evolving templates and wideband measurements provide timing precision at least as good as standard procedures despite the observed frequency drift, however its actual significance will be a matter of debate. This argumentation also applies to this work as ToA/DM uncertainties and RMS of the residuals we obtained reach low, sub-$\mu$s precision which is an improvement when compared to PPTA DR2. A complete and fully restrictive comparison will be possible after obtaining narrow- and wideband results for the next PPTA data release which is already in the process of preparation. 

Nevertheless, observed frequency dependence of the residuals is a noteworthy complication, because wideband timing aims to model all frequency dependent effects in its extraction of a single ToA per observation, thus rendering ad hoc frequency-dependent modeling of the timing residuals obsolete. The solution to this problem is beyond the scope of this work, but is under investigation in several groups.

\subsection{Timing models}
The nominal set of fitted parameters in our timing models included spin period P and spin down rate P1, and five astrometric parameters: right ascension RA and declination DEC, proper motion in both directions PMRA and PMDEC and parallax PX. Binary pulsars were additionally fitted for orbital period PB, projected semi-major axis A1, epoch of ascending node TASC, first and second Laplace parameters (EPS1 and EPS2, respectively). For a few pulsars, proper motions and parallax were poorly constrained so we excluded them from the fitting. In particular, this is the case of pulsars close to the ecliptic such as J1022$+$1001 or J1730-2304. Additionally, first time derivative of orbital period PBDOT, rate of change of periastron and projected semi-major axis (OMDOT and XDOT, respectively) also could not be measured for most pulsars and were excluded (except from models for: J0437$-$4715, J1017$-$7156, J2145$-$0750). For all four pulsars with ELL1H binary model we find both orthometric Shapiro delay parameters H3 and H4. Finally, in case of two pulsars with DDK binary model (J0437$-$3715 and J1713$+$0747) instead of fitting for the aforementioned binary parameters we used epoch and longitude of periastrion (T0 and OM, respectively) and eccentricity of the orbit (E). Tables containing the timing results are presented in Appendix\,\ref{sec:AppB}. Timing model parameters we obtained are consistent with those presented in PPTA DR2 within error limits. The most significant discrepancies were measured for PMRA, PMDEC and PX parameters which is expected because of the much shorter length of our data set (for instance, PX for J2145$-$0750 obtained by us and in PPTA DR2 is 3.4(7) and 1.40(8), respectively). 

\subsection{Noise analysis, timing residuals and DM variations}\label{subsec:noise_residuals}
For the majority of our pulsars EFAC and DMEFAC parameters have values between 1.0 and 1.3 which indicates that the ToA and DM measurements are free of substantial systematic errors (note however, that this is also a function of narrow prior distributions). 

There are two pulsars with excess white noise characterized by a slightly increased DMEFAC (J2124$-$3358 and J2241$-$5236) up to 1.58 and 1.41, respectively. The absolute value of the DMJUMP parameter in most cases is less than 0.0007\,cm$^{-3}$\,pc indicating a proper alignment of profiles comprising the average portrait and a consistent fit for DM from modeling and noise analysis. There are, however, two pulsars (J1824$-$2452A and J2051$-$0827), where DMJUMP goes up to 0.009 and there might be several reasons for that including significant scattering, large DM variations or simply the fact that they have much less data than other pulsars in the data set. Additionally, J2051$-$0827 is a black widow (eclipsing binary) and together with aforementioned ISM effects this can impede obtaining a correct profile alignment. 

Our initial noise models for most pulsars are dominated by large EQUAD values of order of a few hundred nanoseconds. This is due to the fact that EQUADs absorb all of the unmodeled sources of noise including all kinds of red noise but also jitter and issues with modeling the DM variability (usually covered by ECORR parameter which is omitted in wideband methodology).    
We have obtained sub-$\mu$s RMS in 26 pulsars out of which 20 belong to the main PPTA array. Nearly all pulsars (94\%) have median ToA uncertainties lower than $1~\mu$s except two low priority sources which are observed by the PPTA only since recently (J0348$+$0432 and J2150$-$0326). In case of DM uncertainties, our measurements are in the range of (0.043 - 14.24)$\times10^{-4}$\,cm$^{-3}$\,pc. We have obtained DM precision of $10^{-5}$\,cm$^{-3}$\,pc for 10 pulsars and down to $10^{-6}$\,cm$^{-3}$\,pc in the case of two (J0437$-$4715 and J1939$+$2134). This is the level of precision achieved by \cite{Nobleson22} for corresponding pulsars despite the fact that their observations covered 200-500 MHz band where DM measurements can be measured with higher precision because of the larger delays and stronger signal in the low-frequency regime. 

Finally, we would also like to note an observed dependence of the DM on usable bandwidth (cleaned of all spurious channels). For instance, there is a subset of observations, where $f_{\rm ratio} < 3$ because the lower part of the band (below 1400 MHz) was cut out by automatic pipelines due strong RFI. Usually, this resulted in lower or higher estimated DMs when compared to the ones obtained for the whole band even by $\sim 0.005$\,cm$^{-3}$\,pc (see e.g. J0125$-$2327 in Fig.\,\ref{fig:plot0}), which is at least one order of magnitude larger than the typical DM uncertainty obtained in the data set. This effect might be correlated with the observed frequency drift of the residuals reported in Sec. \ref{subsec:subbanded} or to the frequency-dependent DM (explained further in Sec. \ref{subsubsec:J2241_notes}) and will be investigated together in the future work.

\subsection{Pulsars not included in the main PPTA array}\label{subsec:new_pulsars}
Apart from the top priority pulsars observed for nearly two decades, PPTA has also been monitoring 10 new and/or lower priority sources since the start of UWL operation. Their timing potential is currently being investigated and this work may serve as an additional point for their evaluation (Mandow et al. in preparation).

Most of these pulsars (6 out of 10) have low $\rm S/N < 500$, which makes modeling of their profile evolution difficult, however their ToA uncertanties reach precise timing requirements with values between $\sim350$ and $~700$\,ns (apart from J0348$+$0432 and J2150$-$0326 already mentioned in the previous section). The remaining four sources have high S/N and low ToA uncertanties in the range of 129 - 764\,ns. Their models are characterized with good resolution and up to three eigenprofiles, including a fine example of J0125$-$2327 discussed in detail in Sec.\,\ref{subsec:average_portraits} and Sec.\,\ref{subsec:subbanded}. Pulsar J0900$-$3144 is already a part of EPTA with timing baseline of approximately 7 years and ToA precision of $4.27~\mu$s \citep{Desvignes16}, while our work gives $\sigma_{\rm ToA} = 0.612~\mu$s, which indicates that this is a source worth further monitoring. J1933$-$6211 has a profile shape with multipeak leading component strongly evolving with frequency, however we detected only one eigeprofile for it. Further monitoring and collecting more observations would be highly beneficial for a better characterization of this pulsar, especially given that it is a typical binary pulsar with a white dwarf companion \citep{Graikou17} which makes it potentially valuable addition to the PPTA project. J2051$-$0827 is relatively bright ($\rm S/N = 1462$), however it is one of the most difficult sources in our sample to describe. Its pulse shape consists of one peak component strongly affected by scattering and additionally it is an eclipsing binary so modeling its profile evolution and noise analysis are particularly challenging.  
J0348$+$0432 and J2150$-$0326 are the worst two pulsars in our sample in terms of median ToA uncertainties so directly they might not be of much value to the PPTA (although further observations and individual analysis may improve their solutions). Nevertheless, monitoring such sources can also lead to a better understanding of pulsars emission or ISM effects which in turn may help constructing better timing and noise models in general \citep{Kerr20}.

Finally, we also report the measurements of the binary orbital periods for two pulsars, which were not previously published: J0125-2327 and J2051-0827 (Tab.\,\ref{table:1} and Tab.\,\ref{table:6} in Appendix\,\ref{sec:AppB}). 

\subsection{Notes on individuals pulsars}\label{subsec:add_notes}

\subsubsection{J1022$+$1001}\label{subsubsec:J1022_notes}
J1022$+$1001 is an interesting source because of a long-standing controversy regarding its pulse shape instability over timescales ranging from several to several tens of minutes (see \citealp{Padmanabh21} and references therein). There have been various attempts to explain this peculiar behavior, which include suggested polarization calibration errors \citep{Straten13}, strong scintillation coupled with intrinsic profile evolution \citep{Shao16} or effects directly related to the pulsar magnetosphere \citep{Rama03}. Unfortunately, there is still no consensus as these results are often contradicting. 

In our wideband model for this pulsar, a complex shape evolution is evident as the two-peak pulse components change their relative height. Notwithstanding, the reconstructed profiles trace the shape evolution with a good precision (similarly to the model of J0125$-$2327 shown in Fig.\,\ref{fig:J0125_model_obs}).

The spline model for J1022$+$1001 profile evolution required 3 eigenprofiles. They can either correspond to subtle corrections to the intrinsic profile shape changes or to the absorbed ISM/instrumental effects as was mentioned earlier in Sec.\,\ref{subsec:profiles}. It might be tantalising to discern temporal pulse shape variability in the second and/or third eigenprofile, although spline models are supposed to trace only stationary profile changes with frequency. The initial noise model returns a very large EQUAD of 1.4 $\mu$s, which is significantly exceeding recently estimated jitter of $\sim120$ ns for one hour integration \citep{Parthasarathy21} and this might indeed reflect the scatter induced by pulse instability in time (and other noise sources as explained in Sec.\,\ref{subsec:noise_residuals}).

\subsubsection{J2241$-$5236}\label{subsubsec:J2241_notes}

Frequency-dependent DM \citep{Cordes16} is thought to be detectable mostly for bright, high-DM pulsars with little to no profile evolution, which would make J2241$-$5236 a nearly perfect target. If a true nature of phase offsets deviates from $\nu^{-2}$ given by the dispersion law (Eq. \ref{eq:phase}), the DM measured at different parts of the band will vary. In fact, detection of this phenomenon for J2241$-$5236 was recently reported by \citep{Kaur22} based on three days of observations obtained by uGMRT, Murchison Widefield Array (MWA) and Parkes UWL in November 2019. The reported DM changes scale as $\rm \delta DM \sim \nu^{2.5\pm0.1}$ indicating that DM measured at lower frequencies is notably higher then the one measured in the upper part of the band. 

\begin{figure}[ht!]
    \includegraphics[width=\hsize]{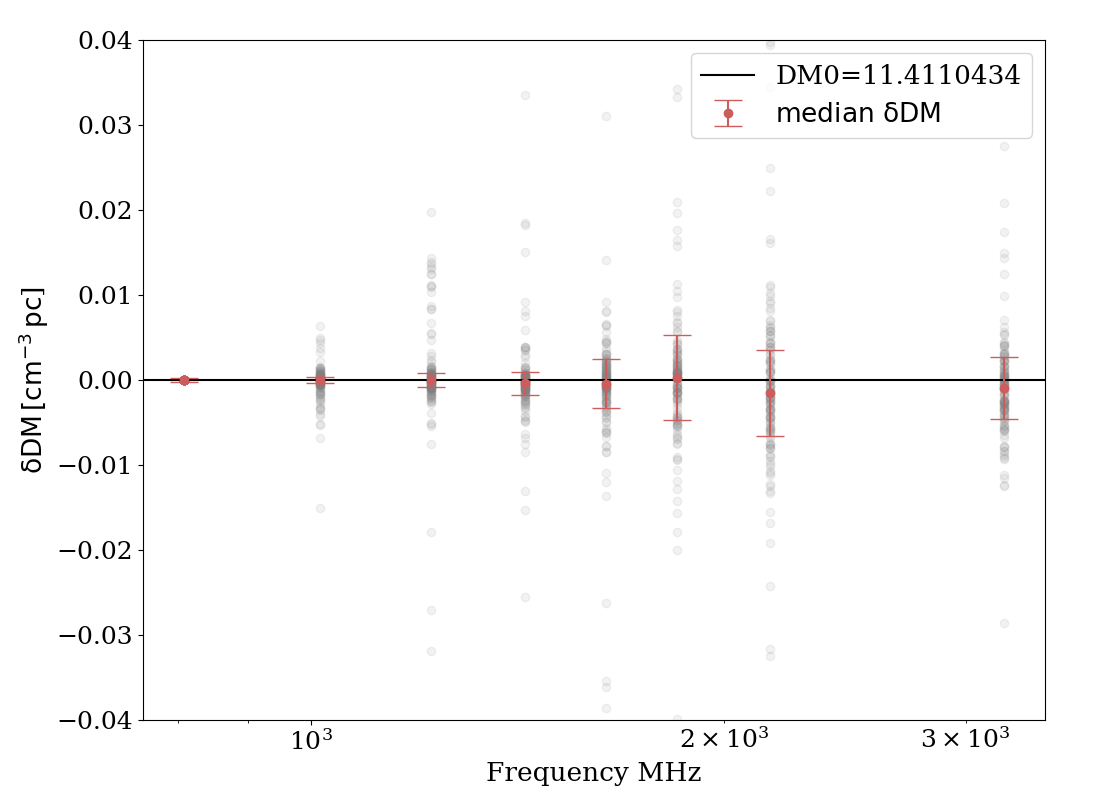}
    \caption{Sub-banded DM measurements as a function of frequency for J2241$-$5236. Grey circles represent each individual measurement, while red points show median values and their uncertainties for each sub-band. The y-axis on the plot was narrowed down to $\rm \delta DM$ in the range of ($-$\,0.1,0.1)\,cm$^{-3}$\,pc for a better visibility omitting a few outlying measurements, however they were included in the calculation of the median value. Nominal DM is shown in the legend of the plot and it corresponds to the black line centered at $\rm \delta DM = 0$.}
    \label{fig:J2241_dm}
\end{figure}

In our wideband analysis, we have initially obtained a similar trend when studying 20-min integrations and data portrait consisting of approximately ten highest S/N observations. The number of detected eigenprofiles strongly depended on the number of used channels (it was either zero or one) despite the overall S/N exceeding 3500. In the final analysis we have used full 1 hour integrations and all available UWL observations which resulted in doubling the S/N of average portraits for most pulsars. Our new model for J2241$-$5236 has two, well resolved eigenprofiles and the sub-banded residuals show a small frequency drift of 0.67 $\mu$s relative to a bin width of 2.14 $\mu$s. 

In Fig.\,\ref{fig:J2241_dm} we show sub-banded DM measurements as a function of frequency. All of our DM measurements now have $\rm \delta DM \sim 0$ with respect to the nominal DM and so any chromatic trend of the DM detected previously was reduced significantly. 

These results indicate that detection of frequency-dependent ISM phenomena may be more difficult than previously expected. Even best candidate pulsars with sharp integrated profile shapes and high stability may exhibit non-negligible profile evolution at small sub-$\mu$s scales, which is challenging both to detect and model. Additionally, we would like to emphasize again that the measured results will be a function of the somewhat arbitrarily determined profile evolution model and disentangling this from actual ISM or other frequency-dependent effects is a non-trivial problem \citep[]{Hassall12}.

\section{Summary and conclusions\label{subsec:summary}}
We have presented wideband timing analysis of the new UWL observations collected under the Parkes Pulsar Timing Array project between November 2018 and March 2022. The main output of our work is a presentation of precise models of profile shape evolution with frequency for 35 pulsars and wideband timing measurements (simultaneously estimated ToAs and DMs). We have also performed an initial noise analysis extended with wideband likelihood including white noise components. Results presented here are the very first demonstration of using methods presented in \cite{Pennucci14} on observations with instantaneous bandwidth larger than 2 GHz (fractional bandwidth of UWL is 1.41), which is soon to become a standard in high precision pulsar timing.  

Precision of our ToA and DM measurements is in the range of 0.005\,$-$\,2.08\,$\mu$s and (0.043\,$-$\,14.24)$\times10^{-4}$\,cm$^{-3}$\,pc, respectively. Comparison of raw ToAs from our work and the previous data release PPTA DR2 yields an increase of the precision by a factor of two, however due to utilization of different receivers, backends and analysis methods this gives only a rough yet promising estimate. A proper comparative analysis will be possible after finalization of two, parallel releases (narrow- and wideband) comprising all available PPTA data extending back to 2004.

In case of two pulsars, J0030$+$0451 and J2129$-$5721, very low S/N above 3000 or 3500 MHz affected the accuracy of the modeled profile evolution. We decided to exclude the uppermost frequencies from our analysis which improved both models, however wideband timing procedures should ultimately be able to resolve such issues in a less brute-force manner. For instance, this could be achieved by adjusting the normalization algorithm so that it does not artificially elevate or overestimate information from low S/N channels.

We have detected a frequency drift of the sub-banded residuals obtained with wideband methodology with an average amplitude of $\sim 1~\mu$s which is below the phase bin resolution of our observations. This might indicate profile alignment issues, however we cannot yet rule out other possible explanations, including timing/noise model systematics, nuances of PCA procedures and the effect of non- or poorly modeled ISM effects, such as scattering. The frequency dependence of the residuals will be investigated further in detail and the results will be presented elsewhere. 

Wideband models presented here will be used for timing analysis of the next full data release (DR3) from PPTA which is now under development. The nominal DR3 data set, despite using frequency evolving templates, will produce sub-banded ToAs of UWL observations in order to combine them with the previous releases in a consistent way. However, we also do intend to produce a wideband DR3 in parallel, where we will reproduce 14-year long data sets with methods presented here and combine it with UWL observations.


\maxdeadcycles=200
\begin{acknowledgments}
The Parkes radio telescope (Murriyang) is part of the Australia Telescope National
Facility which is funded by the Australian Government for operation as a
National Facility managed by CSIRO. MC is supported by the Polish National Science Center through research grant NR 2021/41/N/ST9/01512. SD is the recipient of an Australian Research Council Discovery Early Career Award (DE210101738) funded by the Australian Government.
\end{acknowledgments}

\vspace{5mm}

\software{PulsePortraiture \citep{Pennucci14},  
          Enterprise \citep{Ellis19}, 
          PSRCHIVE \citep{Hotan04},
          Tempo \citep{Nice15},
          Tempo2 \citep{Hobbs06}.
          }

\bibliography{mybib}{}

\begin{thebibliography}{}
\expandafter\ifx\csname natexlab\endcsname\relax\def\natexlab#1{#1}\fi
\providecommand{\url}[1]{\href{#1}{#1}}
\providecommand{\dodoi}[1]{doi:~\href{http://doi.org/#1}{\nolinkurl{#1}}}
\providecommand{\doeprint}[1]{\href{http://ascl.net/#1}{\nolinkurl{http://ascl.net/#1}}}
\providecommand{\doarXiv}[1]{\href{https://arxiv.org/abs/#1}{\nolinkurl{https://arxiv.org/abs/#1}}}

\bibitem[{{Alam} {et~al.}(2021{\natexlab{a}}){Alam}, {Arzoumanian}, {Baker},
  {Blumer}, {Bohler}, {Brazier}, {Brook}, {Burke-Spolaor}, {Caballero},
  {Camuccio}, {Chamberlain}, {Chatterjee}, {Cordes}, {Cornish}, {Crawford},
  {Cromartie}, {Decesar}, {Demorest}, {Dolch}, {Ellis}, {Ferdman}, {Ferrara},
  {Fiore}, {Fonseca}, {Garcia}, {Garver-Daniels}, {Gentile}, {Good},
  {Gusdorff}, {Halmrast}, {Hazboun}, {Islo}, {Jennings}, {Jessup}, {Jones},
  {Kaiser}, {Kaplan}, {Kelley}, {Key}, {Lam}, {Lazio}, {Lorimer}, {Luo},
  {Lynch}, {Madison}, {Maraccini}, {McLaughlin}, {Mingarelli}, {Ng}, {Nguyen},
  {Nice}, {Pennucci}, {Pol}, {Ramette}, {Ransom}, {Ray}, {Shapiro-Albert},
  {Siemens}, {Simon}, {Spiewak}, {Stairs}, {Stinebring}, {Stovall}, {Swiggum},
  {Taylor}, {Tripepi}, {Vallisneri}, {Vigeland}, {Witt}, {Zhu}, \& {Nanograv
  Collaboration}}]{Alam21}
{Alam}, M.~F., {Arzoumanian}, Z., {Baker}, P.~T., {et~al.} 2021{\natexlab{a}},
  \apjs, 252, 5, \dodoi{10.3847/1538-4365/abc6a1}

\bibitem[{{Alam} {et~al.}(2021{\natexlab{b}}){Alam}, {Arzoumanian}, {Baker},
  {Blumer}, {Bohler}, {Brazier}, {Brook}, {Burke-Spolaor}, {Caballero},
  {Camuccio}, {Chamberlain}, {Chatterjee}, {Cordes}, {Cornish}, {Crawford},
  {Cromartie}, {Decesar}, {Demorest}, {Dolch}, {Ellis}, {Ferdman}, {Ferrara},
  {Fiore}, {Fonseca}, {Garcia}, {Garver-Daniels}, {Gentile}, {Good},
  {Gusdorff}, {Halmrast}, {Hazboun}, {Islo}, {Jennings}, {Jessup}, {Jones},
  {Kaiser}, {Kaplan}, {Kelley}, {Key}, {Lam}, {Lazio}, {Lorimer}, {Luo},
  {Lynch}, {Madison}, {Maraccini}, {McLaughlin}, {Mingarelli}, {Ng}, {Nguyen},
  {Nice}, {Pennucci}, {Pol}, {Ramette}, {Ransom}, {Ray}, {Shapiro-Albert},
  {Siemens}, {Simon}, {Spiewak}, {Stairs}, {Stinebring}, {Stovall}, {Swiggum},
  {Taylor}, {Tripepi}, {Vallisneri}, {Vigeland}, {Witt}, {Zhu}, \& {Nanograv
  Collaboration}}]{Alam21b}
---. 2021{\natexlab{b}}, \apjs, 252, 4, \dodoi{10.3847/1538-4365/abc6a0}

\bibitem[{{Amiri} {et~al.}(2021){Amiri}, {Bandura}, {Boyle}, {Brar}, {Cliche},
  {Crowter}, {Cubranic}, {Demorest}, {Denman}, {Dobbs}, {Dong}, {Fandino},
  {Fonseca}, {Good}, {Halpern}, {Hill}, {H{\"o}fer}, {Kaspi}, {Landecker},
  {Leung}, {Lin}, {Luo}, {Masui}, {McKee}, {Mena-Parra}, {Meyers}, {Michilli},
  {Naidu}, {Newburgh}, {Ng}, {Patel}, {Pinsonneault-Marotte}, {Ransom},
  {Renard}, {Scholz}, {Shaw}, {Sikora}, {Stairs}, {Tan}, {Tendulkar},
  {Tretyakov}, {Vanderlinde}, {Wang}, \& {Wang}}]{Amiri21}
{Amiri}, M., {Bandura}, K.~M., {Boyle}, P.~J., {et~al.} 2021, \apjs, 255, 5,
  \dodoi{10.3847/1538-4365/abfdcb}

\bibitem[{{Bulatek} \& {White}(2020)}]{Bulatek20}
{Bulatek}, A., \& {White}, S. 2020, in American Astronomical Society Meeting
  Abstracts, Vol. 235, American Astronomical Society Meeting Abstracts \#235,
  175.17

\bibitem[{{Burke-Spolaor} {et~al.}(2019){Burke-Spolaor}, {Taylor}, {Charisi},
  {Dolch}, {Hazboun}, {Holgado}, {Kelley}, {Lazio}, {Madison}, {McMann},
  {Mingarelli}, {Rasskazov}, {Siemens}, {Simon}, \& {Smith}}]{Burke-Spolaor19}
{Burke-Spolaor}, S., {Taylor}, S.~R., {Charisi}, M., {et~al.} 2019, \aapr, 27,
  5, \dodoi{10.1007/s00159-019-0115-7}

\bibitem[{{CHIME Collaboration} {et~al.}(2022){CHIME Collaboration}, {Amiri},
  {Bandura}, {Boskovic}, {Chen}, {Cliche}, {Deng}, {Denman}, {Dobbs},
  {Fandino}, {Foreman}, {Halpern}, {Hanna}, {Hill}, {Hinshaw}, {H{\"o}fer},
  {Kania}, {Klages}, {Landecker}, {MacEachern}, {Masui}, {Mena-Parra},
  {Milutinovic}, {Mirhosseini}, {Newburgh}, {Nitsche}, {Ordog}, {Pen},
  {Pinsonneault-Marotte}, {Polzin}, {Reda}, {Renard}, {Shaw}, {Siegel},
  {Singh}, {Smegal}, {Tretyakov}, {van Gassen}, {Vanderlinde}, {Wang}, {Wiebe},
  {Willis}, \& {Wulf}}]{Chime22}
{CHIME Collaboration}, {Amiri}, M., {Bandura}, K., {et~al.} 2022, \apjs, 261,
  29, \dodoi{10.3847/1538-4365/ac6fd9}

\bibitem[{{Cordes} {et~al.}(2016){Cordes}, {Shannon}, \&
  {Stinebring}}]{Cordes16}
{Cordes}, J.~M., {Shannon}, R.~M., \& {Stinebring}, D.~R. 2016, \apj, 817, 16,
  \dodoi{10.3847/0004-637X/817/1/16}

\bibitem[{Damour \& Taylor(1992)}]{Damour92}
Damour, T., \& Taylor, J.~H. 1992, Phys. Rev. D, 45, 1840,
  \dodoi{10.1103/PhysRevD.45.1840}

\bibitem[{{Desvignes} {et~al.}(2016){Desvignes}, {Caballero}, {Lentati},
  {Verbiest}, {Champion}, {Stappers}, {Janssen}, {Lazarus}, {Os{\l}owski},
  {Babak}, {Bassa}, {Brem}, {Burgay}, {Cognard}, {Gair}, {Graikou},
  {Guillemot}, {Hessels}, {Jessner}, {Jordan}, {Karuppusamy}, {Kramer},
  {Lassus}, {Lazaridis}, {Lee}, {Liu}, {Lyne}, {McKee}, {Mingarelli},
  {Perrodin}, {Petiteau}, {Possenti}, {Purver}, {Rosado}, {Sanidas}, {Sesana},
  {Shaifullah}, {Smits}, {Taylor}, {Theureau}, {Tiburzi}, {van Haasteren}, \&
  {Vecchio}}]{Desvignes16}
{Desvignes}, G., {Caballero}, R.~N., {Lentati}, L., {et~al.} 2016, \mnras, 458,
  3341, \dodoi{10.1093/mnras/stw483}

\bibitem[{{Ellis} {et~al.}(2019){Ellis}, {Vallisneri}, {Taylor}, \&
  {Baker}}]{Ellis19}
{Ellis}, J.~A., {Vallisneri}, M., {Taylor}, S.~R., \& {Baker}, P.~T. 2019,
  {ENTERPRISE: Enhanced Numerical Toolbox Enabling a Robust PulsaR Inference
  SuitE}.
\newblock \doeprint{1912.015}

\bibitem[{{Fonseca} {et~al.}(2021){Fonseca}, {Cromartie}, {Pennucci}, {Ray},
  {Kirichenko}, {Ransom}, {Demorest}, {Stairs}, {Arzoumanian}, {Guillemot},
  {Parthasarathy}, {Kerr}, {Cognard}, {Baker}, {Blumer}, {Brook}, {DeCesar},
  {Dolch}, {Dong}, {Ferrara}, {Fiore}, {Garver-Daniels}, {Good}, {Jennings},
  {Jones}, {Kaspi}, {Lam}, {Lorimer}, {Luo}, {McEwen}, {McKee}, {McLaughlin},
  {McMann}, {Meyers}, {Naidu}, {Ng}, {Nice}, {Pol}, {Radovan},
  {Shapiro-Albert}, {Tan}, {Tendulkar}, {Swiggum}, {Wahl}, \&
  {Zhu}}]{Fonseca21}
{Fonseca}, E., {Cromartie}, H.~T., {Pennucci}, T.~T., {et~al.} 2021, \apjl,
  915, L12, \dodoi{10.3847/2041-8213/ac03b8}

\bibitem[{{Freire} \& {Wex}(2010)}]{Freire10}
{Freire}, P. C.~C., \& {Wex}, N. 2010, \mnras, 409, 199,
  \dodoi{10.1111/j.1365-2966.2010.17319.x}

\bibitem[{{Goncharov} {et~al.}(2021){Goncharov}, {Reardon}, {Shannon}, {Zhu},
  {Thrane}, {Bailes}, {Bhat}, {Dai}, {Hobbs}, {Kerr}, {Manchester},
  {Os{\l}owski}, {Parthasarathy}, {Russell}, {Spiewak}, {Thyagarajan}, \&
  {Wang}}]{Goncharov21}
{Goncharov}, B., {Reardon}, D.~J., {Shannon}, R.~M., {et~al.} 2021, \mnras,
  502, 478, \dodoi{10.1093/mnras/staa3411}

\bibitem[{{Graikou} {et~al.}(2017){Graikou}, {Verbiest}, {Os{\l}owski},
  {Champion}, {Tauris}, {Jankowski}, \& {Kramer}}]{Graikou17}
{Graikou}, E., {Verbiest}, J.~P.~W., {Os{\l}owski}, S., {et~al.} 2017, \mnras,
  471, 4579, \dodoi{10.1093/mnras/stx1795}

\bibitem[{{Hassall} {et~al.}(2012){Hassall}, {Stappers}, {Hessels}, {Kramer},
  {Alexov}, {Anderson}, {Coenen}, {Karastergiou}, {Keane}, {Kondratiev},
  {Lazaridis}, {van Leeuwen}, {Noutsos}, {Serylak}, {Sobey}, {Verbiest},
  {Weltevrede}, {Zagkouris}, {Fender}, {Wijers}, {B{\"a}hren}, {Bell},
  {Broderick}, {Corbel}, {Daw}, {Dhillon}, {Eisl{\"o}ffel}, {Falcke},
  {Grie{\ss}meier}, {Jonker}, {Law}, {Markoff}, {Miller-Jones}, {Osten}, {Rol},
  {Scaife}, {Scheers}, {Schellart}, {Spreeuw}, {Swinbank}, {ter Veen}, {Wise},
  {Wijnands}, {Wucknitz}, {Zarka}, {Asgekar}, {Bell}, {Bentum}, {Bernardi},
  {Best}, {Bonafede}, {Boonstra}, {Brentjens}, {Brouw}, {Br{\"u}ggen},
  {Butcher}, {Ciardi}, {Garrett}, {Gerbers}, {Gunst}, {van Haarlem}, {Heald},
  {Hoeft}, {Holties}, {de Jong}, {Koopmans}, {Kuniyoshi}, {Kuper}, {Loose},
  {Maat}, {Masters}, {McKean}, {Meulman}, {Mevius}, {Munk}, {Noordam},
  {Orr{\'u}}, {Paas}, {Pandey-Pommier}, {Pandey}, {Pizzo}, {Polatidis},
  {Reich}, {R{\"o}ttgering}, {Sluman}, {Steinmetz}, {Sterks}, {Tagger}, {Tang},
  {Tasse}, {Vermeulen}, {van Weeren}, {Wijnholds}, \& {Yatawatta}}]{Hassall12}
{Hassall}, T.~E., {Stappers}, B.~W., {Hessels}, J.~W.~T., {et~al.} 2012, \aap,
  543, A66, \dodoi{10.1051/0004-6361/201218970}

\bibitem[{{Helfand} {et~al.}(1975){Helfand}, {Manchester}, \&
  {Taylor}}]{Helfand75}
{Helfand}, D.~J., {Manchester}, R.~N., \& {Taylor}, J.~H. 1975, \apj, 198, 661,
  \dodoi{10.1086/153644}

\bibitem[{{Hobbs}(2013)}]{Hobbs2013}
{Hobbs}, G. 2013, Classical and Quantum Gravity, 30, 224007,
  \dodoi{10.1088/0264-9381/30/22/224007}

\bibitem[{{Hobbs} {et~al.}(2019){Hobbs}, {Dai}, {Manchester}, {Shannon},
  {Kerr}, {Lee}, \& {Xu}}]{Hobbs19}
{Hobbs}, G., {Dai}, S., {Manchester}, R.~N., {et~al.} 2019, Research in
  Astronomy and Astrophysics, 19, 020, \dodoi{10.1088/1674-4527/19/2/20}

\bibitem[{{Hobbs} {et~al.}(2020){Hobbs}, {Manchester}, {Dunning}, {Jameson},
  {Roberts}, {George}, {Green}, {Tuthill}, {Toomey}, {Kaczmarek}, {Mader},
  {Marquarding}, {Ahmed}, {Amy}, {Bailes}, {Beresford}, {Bhat}, {Bock},
  {Bourne}, {Bowen}, {Brothers}, {Cameron}, {Carretti}, {Carter}, {Castillo},
  {Chekkala}, {Cheng}, {Chung}, {Craig}, {Dai}, {Dawson}, {Dempsey}, {Doherty},
  {Dong}, {Edwards}, {Ergesh}, {Gao}, {Han}, {Hayman}, {Indermuehle},
  {Jeganathan}, {Johnston}, {Kanoniuk}, {Kesteven}, {Kramer}, {Leach},
  {Mcintyre}, {Moss}, {Os{\l}owski}, {Phillips}, {Pope}, {Preisig}, {Price},
  {Reeves}, {Reilly}, {Reynolds}, {Robishaw}, {Roush}, {Ruckley}, {Sadler},
  {Sarkissian}, {Severs}, {Shannon}, {Smart}, {Smith}, {Smith}, {Sobey},
  {Staveley-Smith}, {Tzioumis}, {van Straten}, {Wang}, {Wen}, \&
  {Whiting}}]{Hobbs20}
{Hobbs}, G., {Manchester}, R.~N., {Dunning}, A., {et~al.} 2020, \pasa, 37,
  e012, \dodoi{10.1017/pasa.2020.2}

\bibitem[{{Hobbs} {et~al.}(2006){Hobbs}, {Edwards}, \& {Manchester}}]{Hobbs06}
{Hobbs}, G.~B., {Edwards}, R.~T., \& {Manchester}, R.~N. 2006, \mnras, 369,
  655, \dodoi{10.1111/j.1365-2966.2006.10302.x}

\bibitem[{{Hotan} {et~al.}(2006){Hotan}, {Bailes}, \& {Ord}}]{Hotan06}
{Hotan}, A.~W., {Bailes}, M., \& {Ord}, S.~M. 2006, \mnras, 369, 1502,
  \dodoi{10.1111/j.1365-2966.2006.10394.x}

\bibitem[{{Hotan} {et~al.}(2004){Hotan}, {van Straten}, \&
  {Manchester}}]{Hotan04}
{Hotan}, A.~W., {van Straten}, W., \& {Manchester}, R.~N. 2004, \pasa, 21, 302,
  \dodoi{10.1071/AS04022}

\bibitem[{{Kaur} {et~al.}(2022){Kaur}, {Ramesh Bhat}, {Dai}, {McSweeney},
  {Shannon}, {Kudale}, \& {van Straten}}]{Kaur22}
{Kaur}, D., {Ramesh Bhat}, N.~D., {Dai}, S., {et~al.} 2022, \apjl, 930, L27,
  \dodoi{10.3847/2041-8213/ac64ff}

\bibitem[{{Kerr} {et~al.}(2020){Kerr}, {Reardon}, {Hobbs}, {Shannon},
  {Manchester}, {Dai}, {Russell}, {Zhang}, {van Straten}, {Os{\l}owski},
  {Parthasarathy}, {Spiewak}, {Bailes}, {Bhat}, {Cameron}, {Coles}, {Dempsey},
  {Deng}, {Goncharov}, {Kaczmarek}, {Keith}, {Lasky}, {Lower}, {Preisig},
  {Sarkissian}, {Toomey}, {Wang}, {Wang}, {Zhang}, \& {Zhu}}]{Kerr20}
{Kerr}, M., {Reardon}, D.~J., {Hobbs}, G., {et~al.} 2020, \pasa, 37, e020,
  \dodoi{10.1017/pasa.2020.11}

\bibitem[{{Kopeikin}(1995)}]{Kopeikin95}
{Kopeikin}, S.~M. 1995, \apjl, 439, L5, \dodoi{10.1086/187731}

\bibitem[{{Kopeikin}(1996)}]{Kopeikin96}
---. 1996, \apjl, 467, L93, \dodoi{10.1086/310201}

\bibitem[{{Kramer} {et~al.}(2016){Kramer}, {Menten}, {Barr}, {Karuppusamy},
  {Kasemann}, {Klein}, {Ros}, {Wieching}, \& {Wucknitz}}]{Kramer16}
{Kramer}, M., {Menten}, K., {Barr}, E.~D., {et~al.} 2016, in MeerKAT Science:
  On the Pathway to the SKA, 3

\bibitem[{{Lange} {et~al.}(2001){Lange}, {Camilo}, {Wex}, {Kramer}, {Backer},
  {Lyne}, \& {Doroshenko}}]{Lange01}
{Lange}, C., {Camilo}, F., {Wex}, N., {et~al.} 2001, \mnras, 326, 274,
  \dodoi{10.1046/j.1365-8711.2001.04606.x}

\bibitem[{{Lee}(2016)}]{Lee16}
{Lee}, K.~J. 2016, in Astronomical Society of the Pacific Conference Series,
  Vol. 502, Frontiers in Radio Astronomy and FAST Early Sciences Symposium
  2015, ed. L.~{Qain} \& D.~{Li}, 19

\bibitem[{{Liu} {et~al.}(2014){Liu}, {Desvignes}, {Cognard}, {Stappers},
  {Verbiest}, {Lee}, {Champion}, {Kramer}, {Freire}, \& {Karuppusamy}}]{Liu14}
{Liu}, K., {Desvignes}, G., {Cognard}, I., {et~al.} 2014, \mnras, 443, 3752,
  \dodoi{10.1093/mnras/stu1420}

\bibitem[{{Lorimer} \& {Kramer}(2004)}]{Lorimer04}
{Lorimer}, D.~R., \& {Kramer}, M. 2004, {Handbook of Pulsar Astronomy}, Vol.~4

\bibitem[{{Luo} {et~al.}(2021){Luo}, {Ransom}, {Demorest}, {Ray}, {Archibald},
  {Kerr}, {Jennings}, {Bachetti}, {van Haasteren}, {Champagne}, {Colen},
  {Phillips}, {Zimmerman}, {Stovall}, {Lam}, \& {Jenet}}]{PINT}
{Luo}, J., {Ransom}, S., {Demorest}, P., {et~al.} 2021, \apj, 911, 45,
  \dodoi{10.3847/1538-4357/abe62f}

\bibitem[{{Manchester} \& {IPTA}(2013)}]{Manchester13}
{Manchester}, R.~N., \& {IPTA}. 2013, Classical and Quantum Gravity, 30,
  224010, \dodoi{10.1088/0264-9381/30/22/224010}

\bibitem[{{McLaughlin}(2013)}]{McLaughlin13}
{McLaughlin}, M.~A. 2013, Classical and Quantum Gravity, 30, 224008,
  \dodoi{10.1088/0264-9381/30/22/224008}

\bibitem[{{Nice} {et~al.}(2015){Nice}, {Demorest}, {Stairs}, {Manchester},
  {Taylor}, {Peters}, {Weisberg}, {Irwin}, {Wex}, \& {Huang}}]{Nice15}
{Nice}, D., {Demorest}, P., {Stairs}, I., {et~al.} 2015, {Tempo: Pulsar timing
  data analysis}.
\newblock \doeprint{1509.002}

\bibitem[{{Nobleson} {et~al.}(2022){Nobleson}, {Agarwal}, {Girgaonkar},
  {Pandian}, {Joshi}, {Krishnakumar}, {Susobhanan}, {Desai}, {Prabu},
  {Bathula}, {Pennucci}, {Banik}, {Bagchi}, {Dhanda Batra}, {Choudhary},
  {Dandapat}, {Dey}, {Gupta}, {Hisano}, {Kato}, {Kharbanda}, {Kikunaga},
  {Kolhe}, {Maan}, {Marmat}, {Arumugam}, {Manoharan}, {Pathak}, {Singha},
  {Surnis}, {Susarla}, \& {Takahashi}}]{Nobleson22}
{Nobleson}, K., {Agarwal}, N., {Girgaonkar}, R., {et~al.} 2022, \mnras, 512,
  1234, \dodoi{10.1093/mnras/stac532}

\bibitem[{{Padmanabh} {et~al.}(2021){Padmanabh}, {Barr}, {Champion},
  {Karuppusamy}, {Kramer}, {Jessner}, \& {Lazarus}}]{Padmanabh21}
{Padmanabh}, P.~V., {Barr}, E.~D., {Champion}, D.~J., {et~al.} 2021, \mnras,
  500, 1178, \dodoi{10.1093/mnras/staa3174}

\bibitem[{{Parthasarathy} {et~al.}(2021){Parthasarathy}, {Bailes}, {Shannon},
  {van Straten}, {Os{\l}owski}, {Johnston}, {Spiewak}, {Reardon}, {Kramer},
  {Venkatraman Krishnan}, {Pennucci}, {Abbate}, {Buchner}, {Camilo},
  {Champion}, {Geyer}, {Hugo}, {Jameson}, {Karastergiou}, {Keith}, \&
  {Serylak}}]{Parthasarathy21}
{Parthasarathy}, A., {Bailes}, M., {Shannon}, R.~M., {et~al.} 2021, \mnras,
  502, 407, \dodoi{10.1093/mnras/stab037}

\bibitem[{Paul {et~al.}(2019)Paul, Susobhanan, Gopakumar, Choudhary, Basu,
  Naidu, Joshi, Pathak, De, M.A., Dey, Bagchi, Surnis, Dhanda, Manoharan,
  Arumugasamy, Susarla, Sanpaarsa, Bethapudi, Desai, Gupta, \& Maan}]{Ashis19}
Paul, A., Susobhanan, A., Gopakumar, A., {et~al.} 2019, in 2019 URSI
  Asia-Pacific Radio Science Conference (AP-RASC), 1--1,
  \dodoi{10.23919/URSIAP-RASC.2019.8738505}

\bibitem[{{Pennucci}(2019)}]{Pennucci19}
{Pennucci}, T.~T. 2019, \apj, 871, 34, \dodoi{10.3847/1538-4357/aaf6ef}

\bibitem[{{Pennucci} {et~al.}(2014){Pennucci}, {Demorest}, \&
  {Ransom}}]{Pennucci14}
{Pennucci}, T.~T., {Demorest}, P.~B., \& {Ransom}, S.~M. 2014, \apj, 790, 93,
  \dodoi{10.1088/0004-637X/790/2/93}

\bibitem[{{Pennucci} {et~al.}(2016){Pennucci}, {Demorest}, \&
  {Ransom}}]{Pennucci16}
---. 2016, {Pulse Portraiture: Pulsar timing}.
\newblock \doeprint{1606.013}

\bibitem[{{Ramachandran} \& {Kramer}(2003)}]{Rama03}
{Ramachandran}, R., \& {Kramer}, M. 2003, \aap, 407, 1085,
  \dodoi{10.1051/0004-6361:20031036}

\bibitem[{{Reardon} {et~al.}(2021){Reardon}, {Shannon}, {Cameron}, {Goncharov},
  {Hobbs}, {Middleton}, {Shamohammadi}, {Thyagarajan}, {Bailes}, {Bhat}, {Dai},
  {Kerr}, {Manchester}, {Russell}, {Spiewak}, {Wang}, \& {Zhu}}]{Reardon21}
{Reardon}, D.~J., {Shannon}, R.~M., {Cameron}, A.~D., {et~al.} 2021, \mnras,
  \dodoi{10.1093/mnras/stab1990}

\bibitem[{{Shannon} \& {Cordes}(2010)}]{Shannon10}
{Shannon}, R.~M., \& {Cordes}, J.~M. 2010, \apj, 725, 1607,
  \dodoi{10.1088/0004-637X/725/2/1607}

\bibitem[{{Shannon} {et~al.}(2014){Shannon}, {Os{\l}owski}, {Dai}, {Bailes},
  {Hobbs}, {Manchester}, {van Straten}, {Raithel}, {Ravi}, {Toomey}, {Bhat},
  {Burke-Spolaor}, {Coles}, {Keith}, {Kerr}, {Levin}, {Sarkissian}, {Wang},
  {Wen}, \& {Zhu}}]{Shannon14}
{Shannon}, R.~M., {Os{\l}owski}, S., {Dai}, S., {et~al.} 2014, \mnras, 443,
  1463, \dodoi{10.1093/mnras/stu1213}

\bibitem[{{Shao} \& {You}(2016)}]{Shao16}
{Shao}, M., \& {You}, X.~P. 2016, Acta Astronomica Sinica, 57, 517

\bibitem[{{Sharma} {et~al.}(2022){Sharma}, {Roy}, {Bhattacharyya}, {Levin},
  {Stappers}, {Pennucci}, {Schult}, {Singh}, \& {Kaninghat}}]{Sharma22}
{Sharma}, S.~S., {Roy}, J., {Bhattacharyya}, B., {et~al.} 2022, arXiv e-prints,
  arXiv:2201.04386.
\newblock \doarXiv{2201.04386}

\bibitem[{{Spiewak} {et~al.}(2022){Spiewak}, {Bailes}, {Miles},
  {Parthasarathy}, {Reardon}, {Shamohammadi}, {Shannon}, {Bhat}, {Buchner},
  {Cameron}, {Camilo}, {Geyer}, {Johnston}, {Karastergiou}, {Keith}, {Kramer},
  {Serylak}, {van Straten}, {Theureau}, \& {Venkatraman Krishnan}}]{Spiewak22}
{Spiewak}, R., {Bailes}, M., {Miles}, M.~T., {et~al.} 2022, arXiv e-prints,
  arXiv:2204.04115.
\newblock \doarXiv{2204.04115}

\bibitem[{{Stovall} {et~al.}(2014){Stovall}, {Lynch}, {Ransom}, {Archibald},
  {Banaszak}, {Biwer}, {Boyles}, {Dartez}, {Day}, {Ford}, {Flanigan}, {Garcia},
  {Hessels}, {Hinojosa}, {Jenet}, {Kaplan}, {Karako-Argaman}, {Kaspi},
  {Kondratiev}, {Leake}, {Lorimer}, {Lunsford}, {Martinez}, {Mata},
  {McLaughlin}, {Roberts}, {Rohr}, {Siemens}, {Stairs}, {van Leeuwen},
  {Walker}, \& {Wells}}]{Stovall14}
{Stovall}, K., {Lynch}, R.~S., {Ransom}, S.~M., {et~al.} 2014, \apj, 791, 67,
  \dodoi{10.1088/0004-637X/791/1/67}

\bibitem[{{Tarafdar} {et~al.}(2022){Tarafdar}, {Nobleson}, {Rana}, {Singha},
  {Krishnakumar}, {Joshi}, {Paladi}, {Kolhe}, {Batra}, {Agarwal}, {Bathula},
  {Dandapat}, {Desai}, {Dey}, {Hisano}, {Ingale}, {Kato}, {Kharbanda},
  {Kikunaga}, {Marmat}, {Pandian}, {Prabu}, {Srivastava}, {Surnis}, {Susarla},
  {Susobhanan}, {Takahashi}, {Arumugam}, {Bagchi}, {Banik}, {De}, {Girgaonkar},
  {Gopakumar}, {Gupta}, {Maan}, {Manoharan}, {Naidu}, \& {Pathak}}]{Tarafdar22}
{Tarafdar}, P., {Nobleson}, K., {Rana}, P., {et~al.} 2022, \pasa, 39, e053,
  \dodoi{10.1017/pasa.2022.46}

\bibitem[{{Vallisneri} {et~al.}(2020){Vallisneri}, {Taylor}, {Simon},
  {Folkner}, {Park}, {Cutler}, {Ellis}, {Lazio}, {Vigeland}, {Aggarwal},
  {Arzoumanian}, {Baker}, {Brazier}, {Brook}, {Burke-Spolaor}, {Chatterjee},
  {Cordes}, {Cornish}, {Crawford}, {Cromartie}, {Crowter}, {DeCesar},
  {Demorest}, {Dolch}, {Ferdman}, {Ferrara}, {Fonseca}, {Garver-Daniels},
  {Gentile}, {Good}, {Hazboun}, {Holgado}, {Huerta}, {Islo}, {Jennings},
  {Jones}, {Jones}, {Kaplan}, {Kelley}, {Key}, {Lam}, {Levin}, {Lorimer},
  {Luo}, {Lynch}, {Madison}, {McLaughlin}, {McWilliams}, {Mingarelli}, {Ng},
  {Nice}, {Pennucci}, {Pol}, {Ransom}, {Ray}, {Siemens}, {Spiewak}, {Stairs},
  {Stinebring}, {Stovall}, {Swiggum}, {van Haasteren}, {Witt}, \&
  {Zhu}}]{Vallisneri20}
{Vallisneri}, M., {Taylor}, S.~R., {Simon}, J., {et~al.} 2020, \apj, 893, 112,
  \dodoi{10.3847/1538-4357/ab7b67}

\bibitem[{{van Straten}(2013)}]{Straten13}
{van Straten}, W. 2013, \apjs, 204, 13, \dodoi{10.1088/0067-0049/204/1/13}

\bibitem[{{van Straten} \& {Bailes}(2011)}]{Straten11}
{van Straten}, W., \& {Bailes}, M. 2011, \pasa, 28, 1, \dodoi{10.1071/AS10021}

\bibitem[{{van Straten} {et~al.}(2001){van Straten}, {Bailes}, {Britton},
  {Kulkarni}, {Anderson}, {Manchester}, \& {Sarkissian}}]{Straten01}
{van Straten}, W., {Bailes}, M., {Britton}, M., {et~al.} 2001, \nat, 412, 158,
  \dodoi{10.1038/35084015}

\end{thebibliography}
\bibliographystyle{aasjournal}

\appendix

\section{Timing residuals and DM measurements\label{sec:AppA}}
Below, we present plots with timing residuals (top panels) and DM variability (bottom panels). For each pulsar, ToA error bars are corrected by the obtained white noise components, namely EFAC and EQUAD, while DM measurements include DMEFAC and DMJUMP. Red and black points in DM plots show the wideband measurements and DMX model, respectively.
If DMX model points are missing in any pulsars plot it means that there is only one DMX bin encompassing the whole timing baseline. This is the case for the least frequently observed pulsars.
\begin{figure}[h]
\centering
    \includegraphics[width=\hsize]{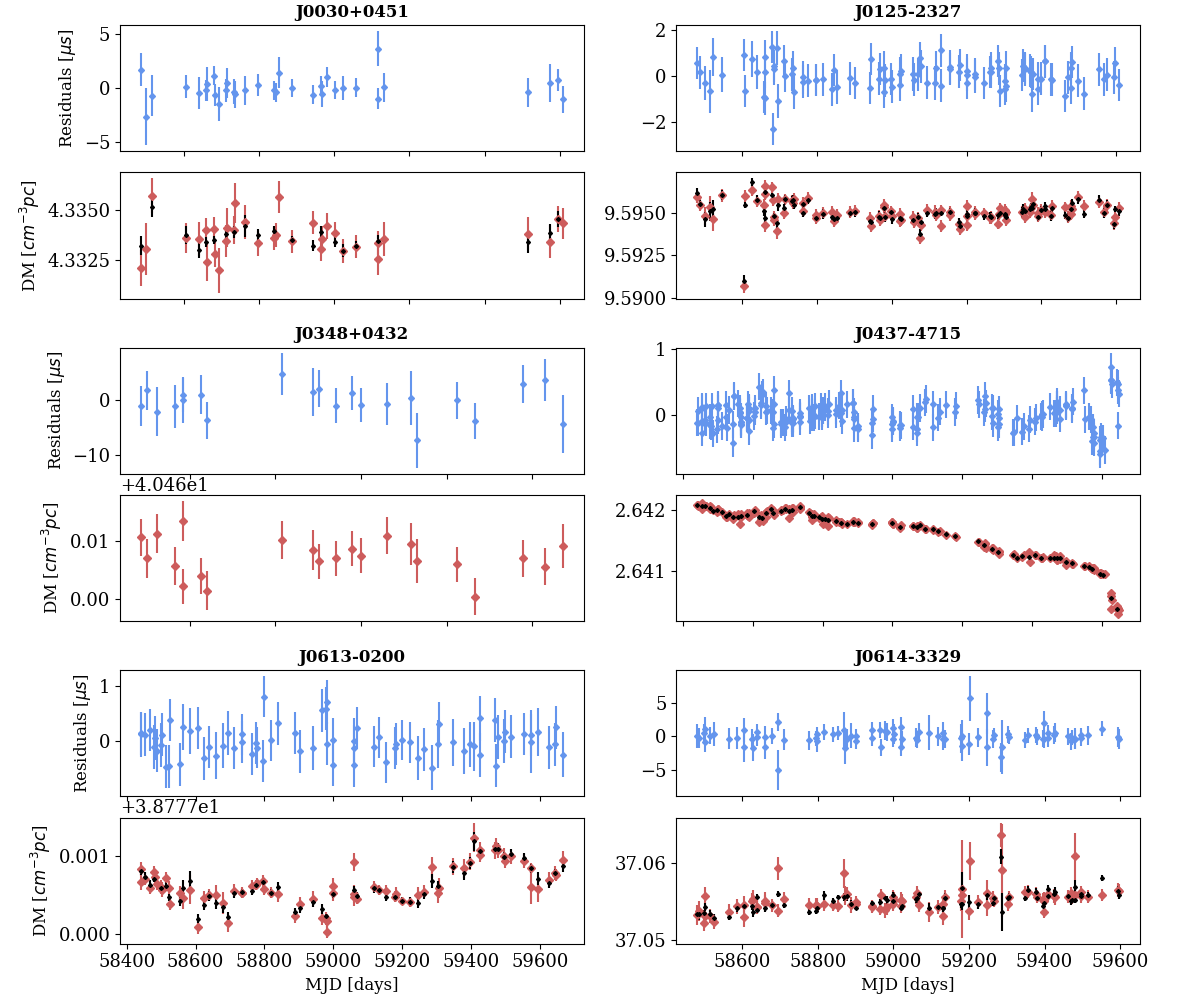}
    \caption{Timing residuals and DM variations for J0030$+$0451, J0125$-$2327, J0348$+$0432, J0437$-$4715, J0613$-$0200 and J0614$-$3329.}
    \label{fig:plot0}
\end{figure}

\begin{figure}[h]
\centering
    \includegraphics[width=\hsize]{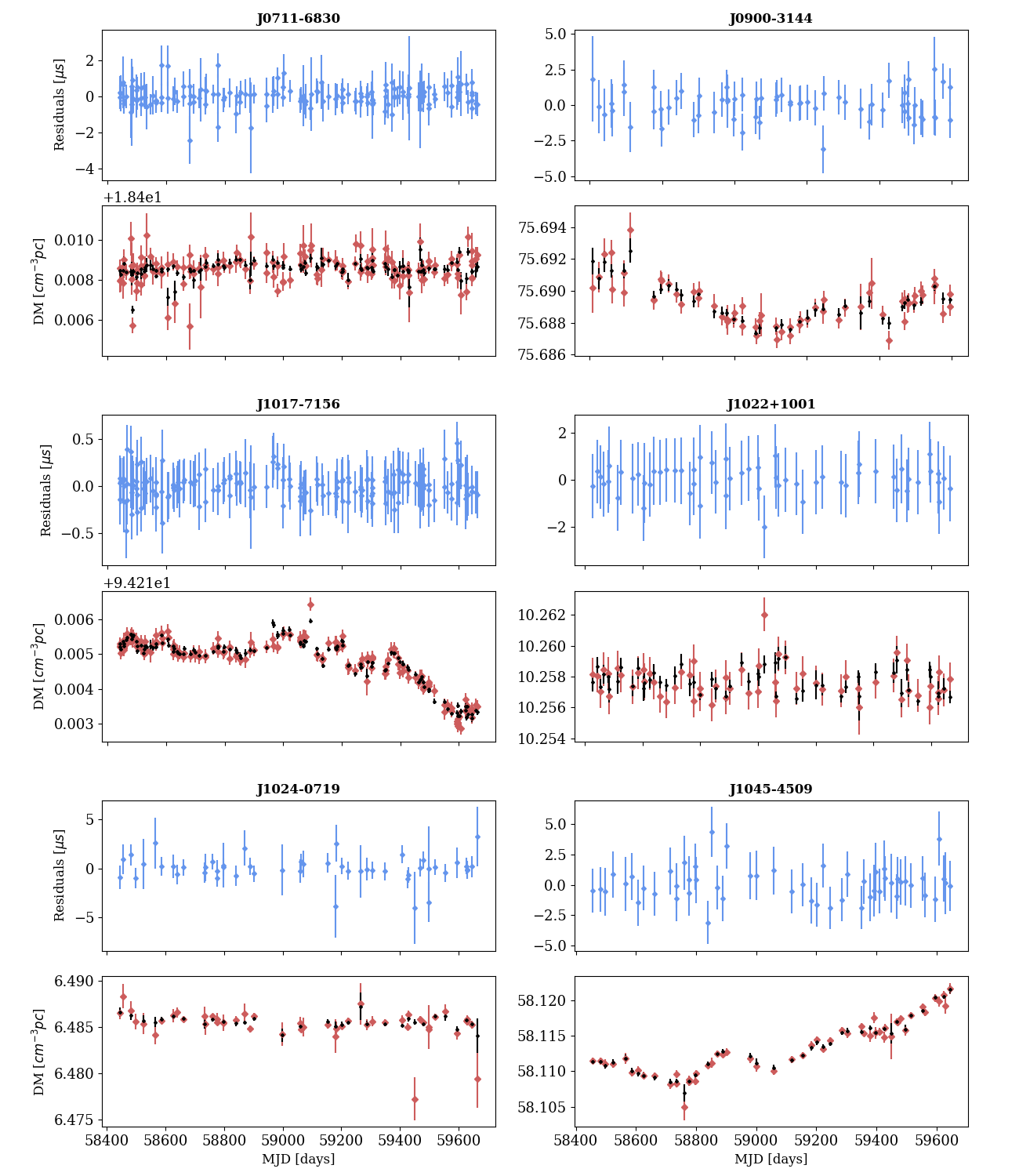}
    \caption{Timing residuals and DM variations for J0711$-$6830, J0900$-$3144, J1017$-$7156, J1022$+$1001,J1024$-$0719 and J1045$-$4509.}
    \label{fig:plot1}
\end{figure}

\begin{figure}[h]
\centering
    \includegraphics[width=\hsize]{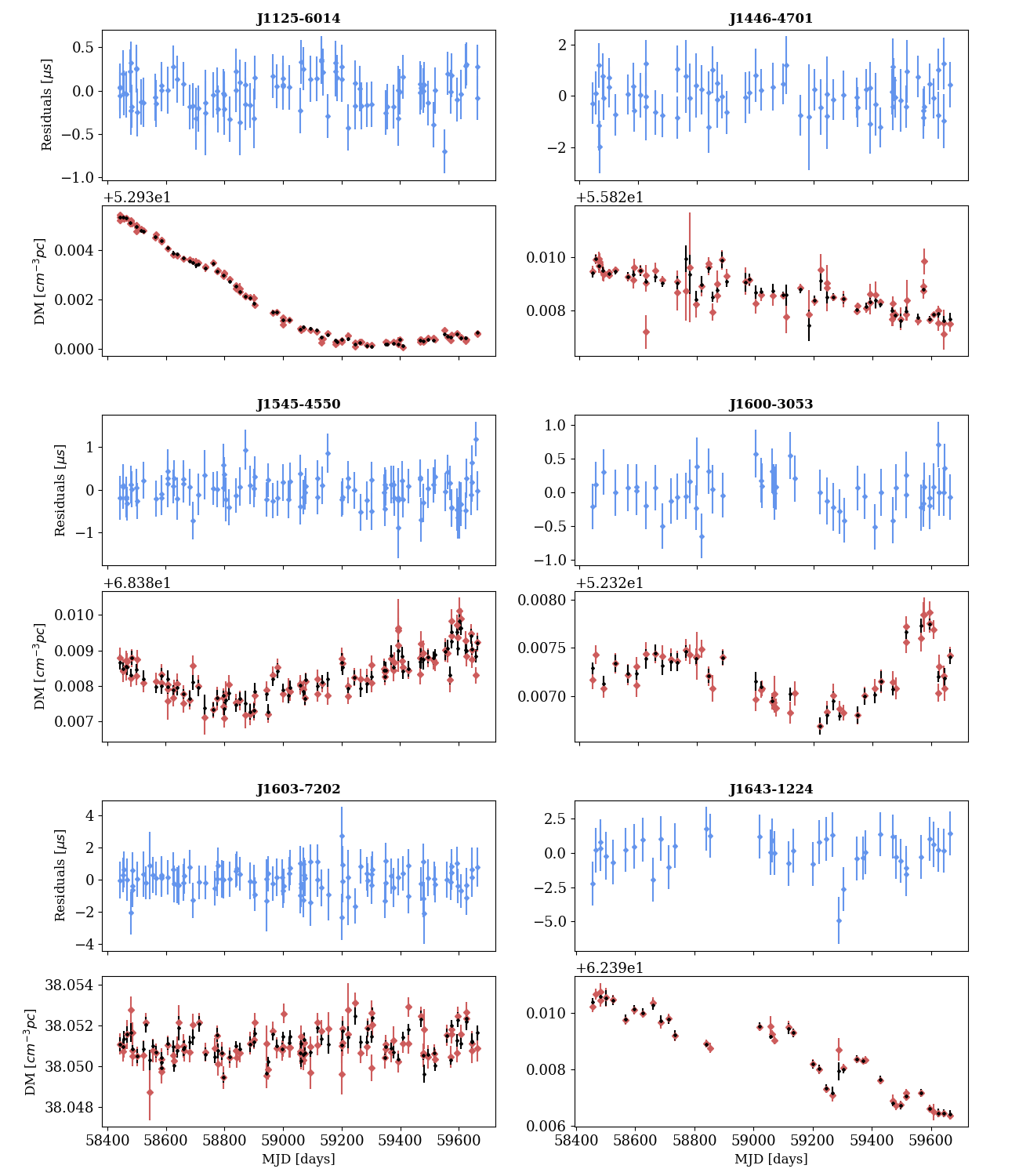}
    \caption{Timing residuals and DM variations for J1125$-$6014, J1446$-$4701,J1545$-$4550, J1600$-$3053, J1603$-$7202 and J1643$-$1224.}
    \label{fig:plot2}
\end{figure}

\begin{figure}[h]
\centering
    \includegraphics[width=\hsize]{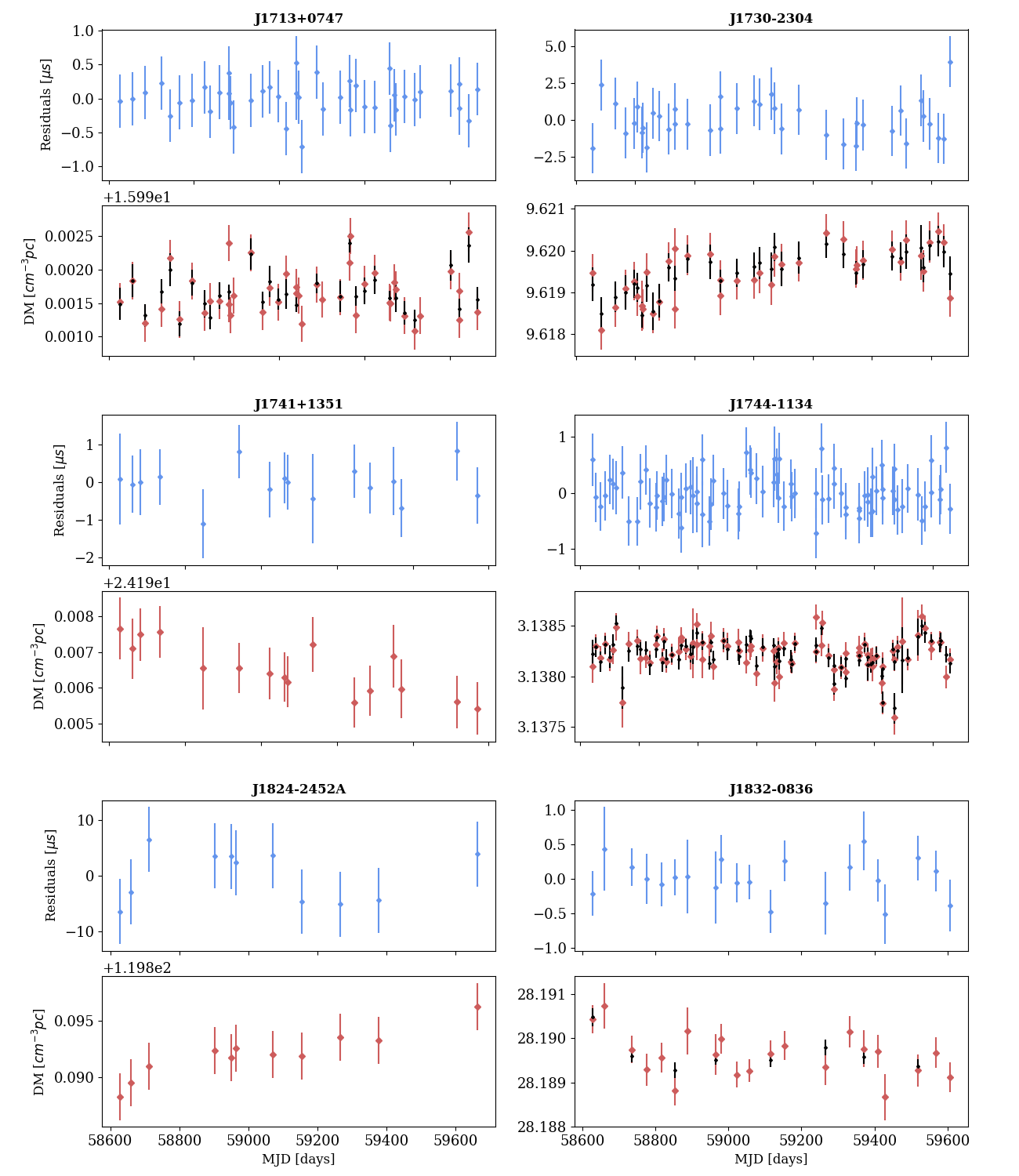}
    \caption{Timing residuals and DM variations for J1713$+$0747, J1730$-$2304, J1741$+$1351, J1744$-$1134, J1824$-$2452A and J1832$-$0836.}
    \label{fig:plot3}
\end{figure}

\begin{figure}[h]
\centering
    \includegraphics[width=\hsize]{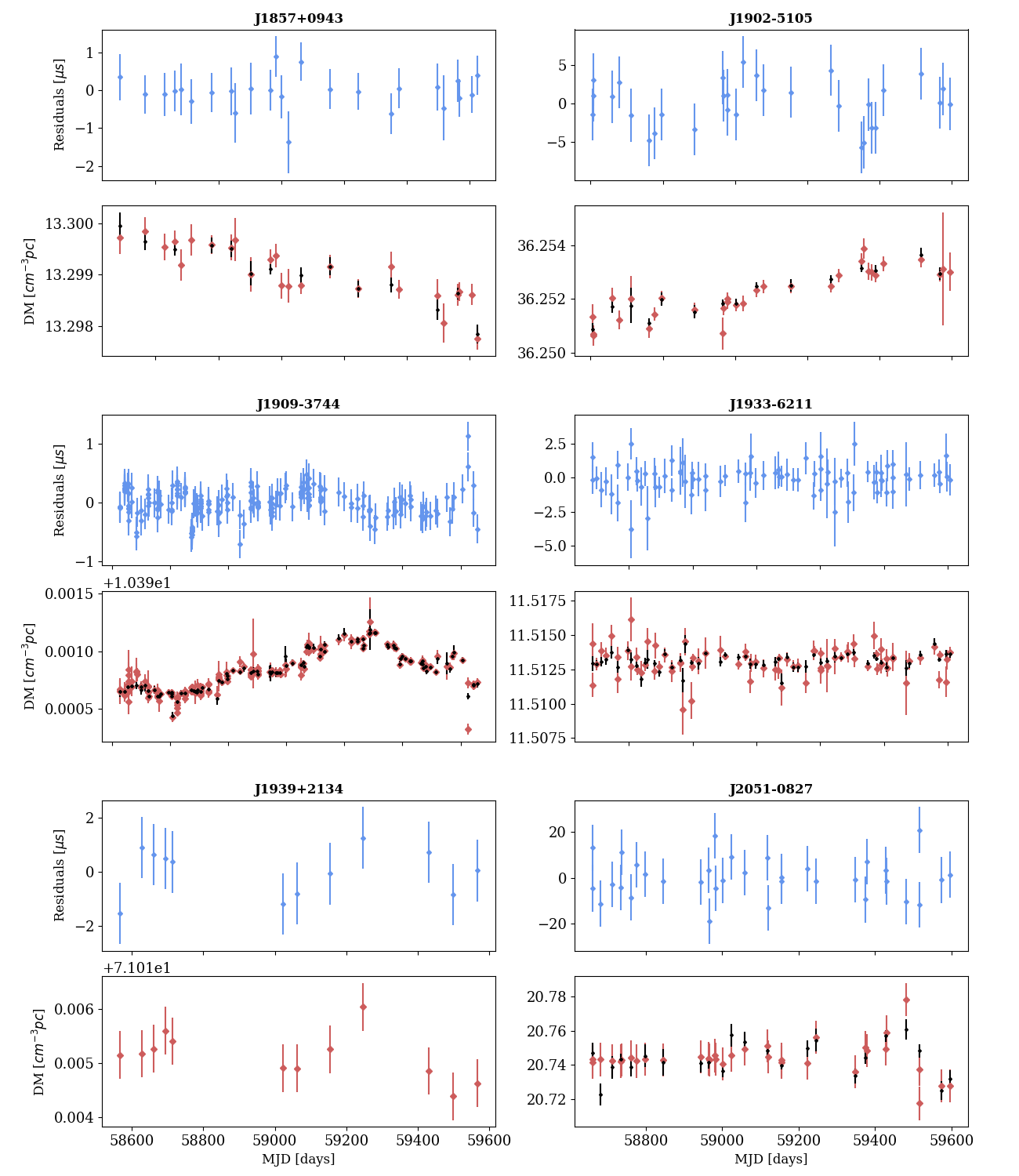}
    \caption{Timing residuals and DM variations for J1857$+$0943, J1902$-$5103, J1909$-$3744, J1933$-$6211, J1939$+$2134 and J2051$-$0827.}
    \label{fig:plot4}
\end{figure}

\begin{figure}[h]
\centering
    \includegraphics[width=\hsize]{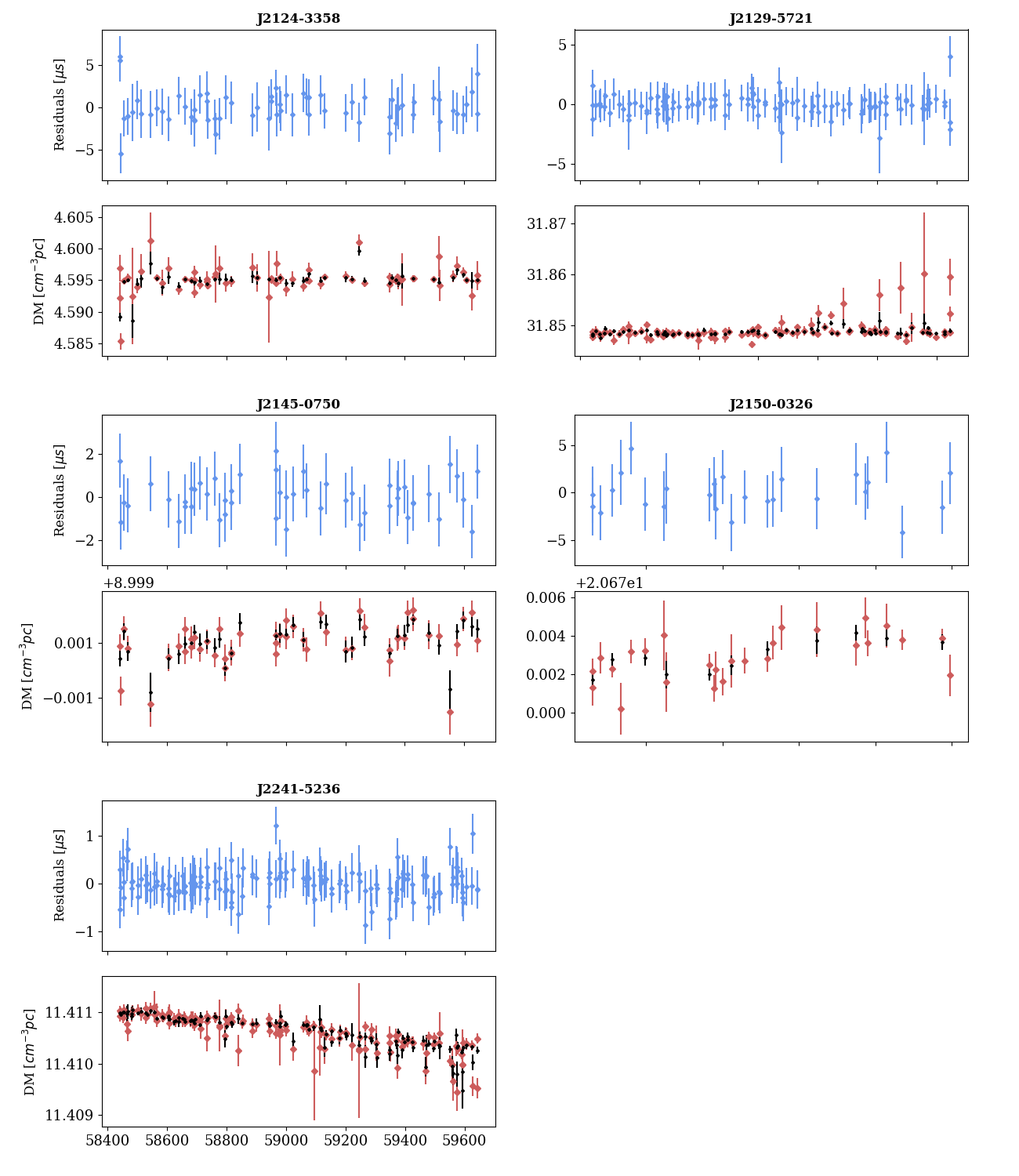}
    \caption{Timing residuals and DM variations for J2124$-$3358, J2129$-$5721, J2145$-$0750, J2150$-$0326 and J2241$-$5236.}
    \label{fig:plot5}
\end{figure}
\newpage

\section{Timing parameters\label{sec:AppB}}
Below we present tables with timing model parameters. The parameters which were used in the fitting are marked in bold.

\movetabledown=65mm
\begin{table*}
\footnotesize
\begin{rotatetable*}
\caption{Timing parameters for J0030$+$0451, J0125$-$2327, J0348$+$0432, J0437$-$4715 and J0613$-$0200.\label{table:1}}
\begin{center}
\begin{tabular}{l l l l l l }
\hline\hline
                                              &      J0030$+$0451      &      J0125$-$2327      &      J0348$+$0432      &      J0437$-$4715      &      J0613$-$0200      \\
\hline
                  MJD range                   &    58486 - 59607     &    58480 - 59607     &    58486 - 59470     &    58441 - 59648     &    58441 - 59665     \\
       Right ascension (RA) [hh:mm:ss]        & \textbf{00:30:27.4234(3)} & \textbf{01:25:1.074328(5)} & \textbf{03:48:43.64127(7)} & \textbf{04:37:16.0445973(9)} & \textbf{06:13:43.977048(3)} \\
        Declination (DEC) [dd:mm:ss]          & \textbf{04:51:39.726(10)} & \textbf{-23:27:8.1349(1)} & \textbf{04:32:11.462(2)} & \textbf{-47:15:9.99983(1)} & \textbf{-02:00:47.3405(1)} \\
     Proper motion in RA [mas $yr^{-1}$]      &       -6.2100        &  \textbf{37.14(8)}   &          --          & \textbf{121.378(8)}  &   \textbf{1.84(4)}   \\
    Proper motion in DEC [mas $yr^{-1}$]      &        0.5000        &   \textbf{10.5(2)}   &          --          &  \textbf{-71.42(1)}  &  \textbf{-10.6(1)}   \\
          Spin frequency [$s^{-1}$]           & \textbf{205.530698975007(4)} & \textbf{272.081088484952(1)} & \textbf{25.560636549613(1)} & \textbf{173.6879478266896(2)} & \textbf{326.6005667292508(8)} \\
    Spin frequency derivative [$s^{-2}$]      & \textbf{-4.296(3)$\cdot10^{-16}$} & \textbf{-1.36332(10)$\cdot10^{-15}$} & \textbf{-1.5764(7)$\cdot10^{-16}$} & \textbf{-1.728459(8)$\cdot10^{-15}$} & \textbf{-1.02301(6)$\cdot10^{-15}$} \\
              DM [$cm^{-3}pc$]                &       4.332798       &       9.598062       &      40.466112       &       2.640302       &      38.779608       \\
               Parallax [mas]                 &        2.7900        &   \textbf{1.4(3)}    &          --          &   \textbf{6.3(3)}    &   \textbf{0.7(2)}    \\
                Binary model                  &          --          &         ELL1         &         ELL1         &         DDK          &         ELL1         \\
             Orbital period [d]               &          --          & \textbf{7.2771996382(8)} & \textbf{1.0242406060(5)$\cdot10^{-1}$} & \textbf{5.74104584368(9)$\cdot10^{-0}$} & \textbf{1.19851255685(5)$\cdot10^{-0}$} \\
      Projected semi-major axis [lt-s]        &          --          & \textbf{4.7297995(1)} & \textbf{0.140988(1)} & \textbf{3.36674563(2)} & \textbf{1.09144391(9)} \\
          Epoch of periastron [MJD]           &          --          &          --          &          --          & \textbf{59048.3754(6)} &          --          \\
        Longitude of periastron [deg]         &          --          &          --          &          --          &   \textbf{1.54(4)}   &          --          \\
            Eccentricity of orbit             &          --          &          --          &          --          & \textbf{0.00001920(1)} &          --          \\
        Epoch of ascending node [MJD]         &          --          & \textbf{59039.36296968(4)} & \textbf{58978.0641405(2)} &          --          & \textbf{59052.42591812(1)} \\
           First Laplace parameter            &          --          & \textbf{0.00000009(6)} & \textbf{-0.00009(3)} &          --          & \textbf{0.0000038(2)} \\
          Second Laplace parameter            &          --          & \textbf{0.00000053(5)} & \textbf{-0.00003(2)} &          --          & \textbf{0.0000032(2)} \\
      Time derivative of orbital period       &          --          &          --          &          --          &          --          &          --          \\
        Rate of advance of periastron         &          --          &          --          &          --          &      0.0137870       &          --          \\
 Rate of change of projected semi-major axis  &          --          &          --          &          --          &          --          &          --          \\
             Sine of inclination              &          --          &          --          &          --          &          --          &          --          \\
        Companion mass [$M_{\odot}$]          &          --          &          --          &          --          &       0.223972       &          --          \\
    Shapiro delay parameter h3 [$\mu s$]      &          --          &          --          &          --          &          --          &          --          \\
    Shapiro delay parameter h4 [$\mu s$]      &          --          &          --          &          --          &          --          &          --          \\
\hline
\end{tabular}
\end{center}
\end{rotatetable*}
\end{table*}

\movetabledown=65mm
\begin{table*}
\footnotesize
\begin{rotatetable*}
\caption{Timing parameters for J0614$-$3329, J0711$-$6830,  J0900$-$3144, J1017$-$7156 and J1022$+$1001.\label{table:2}}
\begin{center}
\begin{tabular}{l l l l l l }
\hline\hline
                                              &      J0614$-$3329      &      J0711$-$6830      &      J0900$-$3144      &      J1017$-$7156      &      J1022$+$1001      \\
\hline
                  MJD range                   &    58481 - 59596     &    58441 - 59664     &    58607 - 59595     &    58441 - 59664     &    58425 - 59665     \\
       Right ascension (RA) [hh:mm:ss]        & \textbf{06:14:10.34823(2)} & \textbf{07:11:54.15385(1)} & \textbf{09:00:43.95220(2)} & \textbf{10:17:51.312124(8)} & \textbf{10:22:57.988(7)} \\
        Declination (DEC) [dd:mm:ss]          & \textbf{-33:29:54.1312(2)} & \textbf{-68:30:47.23742(8)} & \textbf{-31:44:30.8721(2)} & \textbf{-71:56:41.57562(3)} & \textbf{10:01:52.8(3)} \\
     Proper motion in RA [mas $yr^{-1}$]      &        0.5800        &  \textbf{-15.45(7)}  &   \textbf{-1.2(3)}   &  \textbf{-7.36(3)}   &       -16.4260       \\
    Proper motion in DEC [mas $yr^{-1}$]      &       -1.9200        &  \textbf{14.08(7)}   &   \textbf{1.8(3)}    &   \textbf{6.83(3)}   &        1.6741        \\
          Spin frequency [$s^{-1}$]           & \textbf{317.594454701573(4)} & \textbf{182.1172372978466(7)} & \textbf{90.011843174693(1)} & \textbf{427.6219116054646(7)} & \textbf{60.7794488431391(8)} \\
    Spin frequency derivative [$s^{-2}$]      & \textbf{-1.7570(4)$\cdot10^{-15}$} & \textbf{-4.9461(6)$\cdot10^{-16}$} & \textbf{-3.958(1)$\cdot10^{-16}$} & \textbf{-4.0508(6)$\cdot10^{-16}$} & \textbf{-1.6017(8)$\cdot10^{-16}$} \\
              DM [$cm^{-3}pc$]                &      37.049000       &      18.406857       &      75.689026       &      94.223226       &      10.252000       \\
               Parallax [mas]                 &        1.1000        &          --          &        1.8620        &   \textbf{0.9(9)}    &        1.5506        \\
                Binary model                  &         ELL1         &          --          &         ELL1         &        ELL1H         &        ELL1H         \\
             Orbital period [d]               & \textbf{53.58461255(3)} &          --          & \textbf{18.737635770(5)} & \textbf{6.5118987118(2)} & \textbf{7.8051301640(10)} \\
      Projected semi-major axis [lt-s]        & \textbf{27.6387917(6)} &          --          & \textbf{17.2488093(5)} & \textbf{4.83004758(6)} & \textbf{16.7654220(6)} \\
          Epoch of periastron [MJD]           &          --          &          --          &          --          &          --          &          --          \\
        Longitude of periastron [deg]         &          --          &          --          &          --          &          --          &          --          \\
            Eccentricity of orbit             &          --          &          --          &          --          &          --          &          --          \\
        Epoch of ascending node [MJD]         & \textbf{59056.1281453(2)} &          --          & \textbf{59105.63920900(7)} & \textbf{59053.90652703(1)} & \textbf{59048.78549539(4)} \\
           First Laplace parameter            & \textbf{0.00004976(4)} &          --          & \textbf{0.00000986(4)} & \textbf{-0.00007136(2)} & \textbf{0.00009625(7)} \\
          Second Laplace parameter            & \textbf{0.00017382(4)} &          --          & \textbf{0.00000350(4)} & \textbf{0.00012288(2)} & \textbf{-0.00001333(8)} \\
      Time derivative of orbital period       &          --          &          --          &          --          &          --          &          --          \\
        Rate of advance of periastron         &          --          &          --          &          --          &  \textbf{0.016(8)}   &      0.0109124       \\
 Rate of change of projected semi-major axis  &          --          &          --          &          --          &          --          &          --          \\
             Sine of inclination              &          --          &          --          &          --          &          --          &          --          \\
        Companion mass [$M_{\odot}$]          &       0.240000       &          --          &          --          &          --          &          --          \\
    Shapiro delay parameter h3 [$\mu s$]      &          --          &          --          &          --          & \textbf{0.00000011(4)} & \textbf{0.0000011(4)} \\
    Shapiro delay parameter h4 [$\mu s$]      &          --          &          --          &          --          & \textbf{0.00000010(5)} & \textbf{0.0000004(5)} \\
\hline
\end{tabular}
\end{center}
\end{rotatetable*}
\end{table*}

\movetabledown=65mm
\begin{table*}
\footnotesize
\begin{rotatetable*}
\caption{Timing parameters for J1024$-$0719, J1045$-$4509, J1125$-$6014, J1446$-$4701 and J1545$-$4550.\label{table:3}}
\begin{center}
\begin{tabular}{l l l l l l }
\hline\hline
                                              &      J1024$-$0719      &      J1045$-$4509      &      J1125$-$6014      &      J1446$-$4701      &      J1545$-$4550      \\
\hline
                  MJD range                   &    58443 - 59665     &    58454 - 59645     &    58442 - 59665     &    58443 - 59665     &    58442 - 59665     \\
       Right ascension (RA) [hh:mm:ss]        & \textbf{10:24:38.64907(2)} & \textbf{10:45:50.17978(3)} & \textbf{11:25:55.242909(3)} & \textbf{14:46:35.70981(1)} & \textbf{15:45:55.945423(4)} \\
        Declination (DEC) [dd:mm:ss]          & \textbf{-07:19:19.9693(6)} & \textbf{-45:09:54.0569(3)} & \textbf{-60:14:6.80611(3)} & \textbf{-47:01:26.7964(2)} & \textbf{-45:50:37.50900(9)} \\
     Proper motion in RA [mas $yr^{-1}$]      &  \textbf{-35.3(3)}   &   \textbf{-5.9(3)}   &  \textbf{11.13(3)}   &  \textbf{-4.20(10)}  &  \textbf{-0.35(4)}   \\
    Proper motion in DEC [mas $yr^{-1}$]      &  \textbf{-48.1(7)}   &   \textbf{4.4(3)}    &  \textbf{-13.04(3)}  &   \textbf{-2.8(2)}   &   \textbf{2.39(8)}   \\
          Spin frequency [$s^{-1}$]           & \textbf{193.715686208170(5)} & \textbf{133.793151505030(2)} & \textbf{380.1730997159584(5)} & \textbf{455.644022907970(2)} & \textbf{279.6977022449402(6)} \\
    Spin frequency derivative [$s^{-2}$]      & \textbf{-6.971(1)$\cdot10^{-16}$} & \textbf{-3.165(2)$\cdot10^{-16}$} & \textbf{-5.3938(3)$\cdot10^{-16}$} & \textbf{-2.0362(2)$\cdot10^{-15}$} & \textbf{-4.10347(4)$\cdot10^{-15}$} \\
              DM [$cm^{-3}pc$]                &       6.480135       &      58.150964       &      52.934539       &      55.830112       &      68.392666       \\
               Parallax [mas]                 &   \textbf{1.1(8)}    &        3.3416        &   \textbf{1.7(3)}    &   \textbf{0.4(5)}    &   \textbf{0.8(2)}    \\
                Binary model                  &          --          &         ELL1         &         ELL1         &         ELL1         &        ELL1H         \\
             Orbital period [d]               &          --          & \textbf{4.083529193(1)} & \textbf{8.7526035149(2)} & \textbf{2.7766607312(9)$\cdot10^{-1}$} & \textbf{6.2030648297(3)} \\
      Projected semi-major axis [lt-s]        &          --          & \textbf{3.0151308(5)} & \textbf{8.3391936(5)} & \textbf{0.0640121(2)} & \textbf{3.84690438(9)} \\
          Epoch of periastron [MJD]           &          --          &          --          &          --          &          --          &          --          \\
        Longitude of periastron [deg]         &          --          &          --          &          --          &          --          &          --          \\
            Eccentricity of orbit             &          --          &          --          &          --          &          --          &          --          \\
        Epoch of ascending node [MJD]         &          --          & \textbf{59049.0111262(1)} & \textbf{59053.335201025(8)} & \textbf{59053.9339604(1)} & \textbf{59056.49624717(2)} \\
           First Laplace parameter            &          --          & \textbf{-0.0000205(4)} & \textbf{-0.00000002(3)} & \textbf{0.000003(6)} & \textbf{-0.00000871(5)} \\
          Second Laplace parameter            &          --          & \textbf{-0.0000104(4)} & \textbf{-0.00000064(1)} & \textbf{0.000010(6)} & \textbf{-0.00000968(4)} \\
      Time derivative of orbital period       &          --          &          --          &          --          &          --          &          --          \\
        Rate of advance of periastron         &          --          &          --          &          --          &          --          &          --          \\
 Rate of change of projected semi-major axis  &          --          &          --          &          --          &          --          &          --          \\
             Sine of inclination              &          --          &          --          &  \textbf{0.977(7)}   &          --          &          --          \\
        Companion mass [$M_{\odot}$]          &          --          &          --          &   \textbf{0.33(5)}   &          --          &          --          \\
    Shapiro delay parameter h3 [$\mu s$]      &          --          &          --          &          --          &          --          & \textbf{0.00000007(6)} \\
    Shapiro delay parameter h4 [$\mu s$]      &          --          &          --          &          --          &          --          & \textbf{-0.00000010(9)} \\
\hline
\end{tabular}
\end{center}
\end{rotatetable*}
\end{table*}

\movetabledown=65mm
\begin{table*}
\footnotesize
\begin{rotatetable*}
\caption{Parameter files.\label{table:4}}
\begin{center}
\begin{tabular}{l l l l l l }
\hline\hline
                                              &      J1600-3053      &      J1603-7202      &      J1643-1224      &      J1713+0747      &      J1730-2304      \\
\hline
                  MJD range                   &    58443 - 59665     &    58443 - 59664     &    58454 - 59664     &    58426 - 59265     &    58455 - 59664     \\
       Right ascension (RA) [hh:mm:ss]        & \textbf{16:00:51.902531(5)} & \textbf{16:03:35.67088(3)} & \textbf{16:43:38.16602(2)} & \textbf{17:13:49.536688(5)} & \textbf{17:30:21.6849(3)} \\
        Declination (DEC) [dd:mm:ss]          & \textbf{-30:53:49.4528(2)} & \textbf{-72:02:32.8216(2)} & \textbf{-12:24:58.633(1)} & \textbf{07:47:37.4514(1)} & \textbf{-23:04:31.12(7)} \\
     Proper motion in RA [mas $yr^{-1}$]      &  \textbf{-0.77(6)}   &   \textbf{-2.4(1)}   &   \textbf{5.9(2)}    &   \textbf{4.7(1)}    &       20.0246        \\
    Proper motion in DEC [mas $yr^{-1}$]      &   \textbf{-6.8(2)}   &   \textbf{-7.2(2)}   &   \textbf{1.9(10)}   &   \textbf{-3.9(4)}   &       -4.9776        \\
          Spin frequency [$s^{-1}$]           & \textbf{277.9377110420382(7)} & \textbf{67.3765821486238(5)} & \textbf{216.373340194411(2)} & \textbf{218.811843674173(2)} & \textbf{123.110288948911(1)} \\
    Spin frequency derivative [$s^{-2}$]      & \textbf{-7.3383(5)$\cdot10^{-16}$} & \textbf{-7.092(4)$\cdot10^{-17}$} & \textbf{-8.645(2)$\cdot10^{-16}$} & \textbf{-4.086(3)$\cdot10^{-16}$} & \textbf{-3.060(1)$\cdot10^{-16}$} \\
              DM [$cm^{-3}pc$]                &      52.328259       &      38.042070       &      62.301899       &      15.917131       &       9.622926       \\
               Parallax [mas]                 &   \textbf{1.0(2)}    &        0.7019        &   \textbf{2.6(8)}    &   \textbf{0.0(4)}    &   \textbf{1.7(8)}    \\
                Binary model                  &        ELL1H         &         ELL1         &          DD          &         DDK          &          --          \\
             Orbital period [d]               & \textbf{14.3484575522(8)} & \textbf{6.3086295734(7)} & \textbf{147.0173956(1)} & \textbf{67.82512992(2)} &          --          \\
      Projected semi-major axis [lt-s]        & \textbf{8.8016553(1)} & \textbf{6.8806683(3)} & \textbf{25.0725735(5)} & \textbf{32.3424220(1)} &          --          \\
          Epoch of periastron [MJD]           &          --          &          --          & \textbf{58987.082(2)} & \textbf{58847.926(1)} &          --          \\
        Longitude of periastron [deg]         &          --          &          --          & \textbf{321.849(4)}  & \textbf{176.240(6)}  &          --          \\
            Eccentricity of orbit             &          --          &          --          & \textbf{0.00050576(3)} & \textbf{0.000074955(9)} &          --          \\
        Epoch of ascending node [MJD]         & \textbf{59056.37152772(2)} & \textbf{59050.18353655(4)} &          --          &          --          &          --          \\
           First Laplace parameter            & \textbf{-0.00000546(2)} & \textbf{0.00000160(8)} &          --          &          --          &          --          \\
          Second Laplace parameter            & \textbf{-0.00017365(2)} & \textbf{-0.00000925(7)} &          --          &          --          &          --          \\
      Time derivative of orbital period       &          --          &          --          &          --          &          --          &          --          \\
        Rate of advance of periastron         &  \textbf{-0.007(8)}  &          --          &          --          &          --          &          --          \\
 Rate of change of projected semi-major axis  &          --          &          --          &  \textbf{-0.04(1)}   &          --          &          --          \\
             Sine of inclination              &          --          &          --          &          --          &          --          &          --          \\
        Companion mass [$M_{\odot}$]          &          --          &          --          &          --          &       0.538032       &          --          \\
    Shapiro delay parameter h3 [$\mu s$]      & \textbf{0.00000032(7)} &          --          &          --          &          --          &          --          \\
    Shapiro delay parameter h4 [$\mu s$]      & \textbf{0.00000011(10)} &          --          &          --          &          --          &          --          \\
\hline
\end{tabular}
\end{center}
\end{rotatetable*}
\end{table*}

\movetabledown=65mm
\begin{table*}
\footnotesize
\begin{rotatetable*}
\caption{Timing parameters for J1741$+$1351, J1744$-$1134, J1824$-$2452A, J1832$-$0836 and J1857$+$0943.\label{table:5}}
\begin{center}
\begin{tabular}{l l l l l l }
\hline\hline
                                              &      J1741$+$1351      &      J1744$-$1134      &     J1824$-$2452A      &      J1832$-$0836      &      J1857$+$0943      \\
\hline
                  MJD range                   &    58627 - 59571     &    58442 - 59658     &    58627 - 59664     &    58627 - 59607     &    58486 - 59624     \\
       Right ascension (RA) [hh:mm:ss]        & \textbf{17:41:31.13983(1)} & \textbf{17:44:29.421749(3)} & \textbf{18:24:32.0076(2)} & \textbf{18:32:27.589563(5)} & \textbf{18:57:36.388636(7)} \\
        Declination (DEC) [dd:mm:ss]          & \textbf{13:51:44.0624(3)} & \textbf{-11:34:54.7976(2)} & \textbf{-24:52:10.94(5)} & \textbf{-08:36:55.1437(3)} & \textbf{09:43:17.1470(2)} \\
     Proper motion in RA [mas $yr^{-1}$]      &       -8.7609        &  \textbf{18.86(4)}   &       -0.2339        &  \textbf{-8.13(8)}   &   \textbf{-2.6(1)}   \\
    Proper motion in DEC [mas $yr^{-1}$]      &       -6.5447        &   \textbf{-9.5(2)}   &       -8.9459        &  \textbf{-21.2(4)}   &   \textbf{-5.9(2)}   \\
          Spin frequency [$s^{-1}$]           & \textbf{266.869166434737(3)} & \textbf{245.4261233068814(5)} & \textbf{327.40552589863(2)} & \textbf{367.767120938621(2)} & \textbf{186.494081052219(1)} \\
    Spin frequency derivative [$s^{-2}$]      & \textbf{-2.1524(2)$\cdot10^{-15}$} & \textbf{-5.3818(4)$\cdot10^{-16}$} & \textbf{-1.73531(2)$\cdot10^{-13}$} & \textbf{-1.1194(2)$\cdot10^{-15}$} & \textbf{-6.2038(8)$\cdot10^{-16}$} \\
              DM [$cm^{-3}pc$]                &      24.198581       &       3.139444       &      119.902567      &      28.195970       &      13.298525       \\
               Parallax [mas]                 &        0.4586        &   \textbf{3.1(2)}    &       -0.2430        &   \textbf{1.1(3)}    &   \textbf{1.6(5)}    \\
                Binary model                  &         ELL1         &          --          &          --          &          --          &         ELL1         \\
             Orbital period [d]               & \textbf{16.335347843(6)} &          --          &          --          &          --          &  12.32717119165031   \\
      Projected semi-major axis [lt-s]        & \textbf{11.0033142(4)} &          --          &          --          &          --          &     9.230780241      \\
          Epoch of periastron [MJD]           &          --          &          --          &          --          &          --          &          --          \\
        Longitude of periastron [deg]         &          --          &          --          &          --          &          --          &          --          \\
            Eccentricity of orbit             &          --          &          --          &          --          &          --          &          --          \\
        Epoch of ascending node [MJD]         & \textbf{59102.65959876(9)} &          --          &          --          &          --          & \textbf{59058.66451258(5)} \\
           First Laplace parameter            & \textbf{-0.00000399(8)} &          --          &          --          &          --          &    -0.0000215720     \\
          Second Laplace parameter            & \textbf{-0.00000896(7)} &          --          &          --          &          --          &     0.0000024568     \\
      Time derivative of orbital period       &          --          &          --          &          --          &          --          &          --          \\
        Rate of advance of periastron         &          --          &          --          &          --          &          --          &          --          \\
 Rate of change of projected semi-major axis  &          --          &          --          &          --          &          --          &          --          \\
             Sine of inclination              &       0.941114       &          --          &          --          &          --          &  \textbf{0.9990(4)}  \\
        Companion mass [$M_{\odot}$]          &       0.254703       &          --          &          --          &          --          &       0.253681       \\
    Shapiro delay parameter h3 [$\mu s$]      &          --          &          --          &          --          &          --          &          --          \\
    Shapiro delay parameter h4 [$\mu s$]      &          --          &          --          &          --          &          --          &          --          \\
\hline
\end{tabular}
\end{center}
\end{rotatetable*}
\end{table*}

\movetabledown=65mm
\begin{table*}
\footnotesize
\begin{rotatetable*}
\caption{Timing parameters for J1902$-$5105, J1909$-$3744, J1933$-$6211, J1939$+$2134 and J2051$-$0827.\label{table:6}}
\begin{center}
\begin{tabular}{l l l l l l }
\hline\hline
                                              &      J1902-5105      &      J1909-3744      &      J1933-6211      &      J1939+2134      &      J2051-0827      \\
\hline
                  MJD range                   &    58604 - 59595     &    58427 - 59658     &    58485 - 59607     &    58566 - 59567     &    58659 - 59596     \\
       Right ascension (RA) [hh:mm:ss]        & \textbf{19:02:2.84424(4)} & \textbf{19:09:47.424713(1)} & \textbf{19:33:32.41353(2)} & \textbf{19:39:38.56131(2)} & \textbf{20:51:7.5244(4)} \\
        Declination (DEC) [dd:mm:ss]          & \textbf{-51:05:57.0563(9)} & \textbf{-37:44:14.91222(5)} & \textbf{-62:11:46.6962(2)} & \textbf{21:34:59.1205(3)} & \textbf{-08:27:37.76(2)} \\
     Proper motion in RA [mas $yr^{-1}$]      &       -3.0375        &  \textbf{-9.49(1)}   &   \textbf{-5.4(2)}   &        0.0610        &        6.4920        \\
    Proper motion in DEC [mas $yr^{-1}$]      &       -11.1896       &  \textbf{-35.72(5)}  &   \textbf{11.2(2)}   &       -0.4044        &        0.8409        \\
          Spin frequency [$s^{-1}$]           & \textbf{573.92104402934(2)} & \textbf{339.3156919155664(2)} & \textbf{282.212317830028(2)} & \textbf{641.92821931762(1)} & \textbf{221.79628690412(4)} \\
    Spin frequency derivative [$s^{-2}$]      & \textbf{-3.035(1)$\cdot10^{-15}$} & \textbf{-1.61469(2)$\cdot10^{-15}$} & \textbf{-3.078(2)$\cdot10^{-16}$} & \textbf{-4.33064(7)$\cdot10^{-14}$} & \textbf{-6.36(4)$\cdot10^{-16}$} \\
              DM [$cm^{-3}pc$]                &      36.250000       &      10.391734       &      11.499000       &      71.014000       &      20.729900       \\
               Parallax [mas]                 &          --          &   \textbf{1.26(5)}   &          --          &        0.2368        &          --          \\
                Binary model                  &         ELL1         &         ELL1         &         ELL1         &          --          &         ELL1         \\
             Orbital period [d]               & \textbf{2.011803739(1)} & \textbf{1.53344945354(2)$\cdot10^{-0}$} & \textbf{12.819406522(2)} &          --          & \textbf{0.0991102480(6)} \\
      Projected semi-major axis [lt-s]        &     1.901956478      & \textbf{1.89799100(10)} & \textbf{12.2815798(3)} &          --          & \textbf{0.045075(4)} \\
          Epoch of periastron [MJD]           &          --          &          --          &          --          &          --          &          --          \\
        Longitude of periastron [deg]         &          --          &          --          &          --          &          --          &          --          \\
            Eccentricity of orbit             &          --          &          --          &          --          &          --          &          --          \\
        Epoch of ascending node [MJD]         & \textbf{59099.3814765(2)} & \textbf{59042.266169044(4)} & \textbf{59051.25497668(4)} &          --          & \textbf{59127.024438(2)} \\
           First Laplace parameter            &     0.0000000000     & \textbf{0.00000007(5)} & \textbf{0.00000135(4)} &          --          &     0.0000000000     \\
          Second Laplace parameter            &     0.0000000000     & \textbf{-0.00000007(3)} & \textbf{-0.00000029(4)} &          --          &     0.0000000000     \\
      Time derivative of orbital period       &          --          &          --          &          --          &          --          &          --          \\
        Rate of advance of periastron         &          --          &          --          &          --          &          --          &          --          \\
 Rate of change of projected semi-major axis  &          --          &          --          &          --          &          --          &          --          \\
             Sine of inclination              &          --          &  \textbf{0.9983(3)}  &          --          &          --          &          --          \\
        Companion mass [$M_{\odot}$]          &          --          &  \textbf{0.202(6)}   &          --          &          --          &          --          \\
    Shapiro delay parameter h3 [$\mu s$]      &          --          &          --          &          --          &          --          &          --          \\
    Shapiro delay parameter h4 [$\mu s$]      &          --          &          --          &          --          &          --          &          --          \\
\hline
\end{tabular}
\end{center}
\end{rotatetable*}
\end{table*}

\movetabledown=65mm
\begin{table*}
\footnotesize
\begin{rotatetable*}
\caption{Timing parameters J2124$-$3358, J2129$-$5721, J2145$-$0750, J2150$-$0326 and J2241$-$5236.\label{table:7}}
\begin{center}
\begin{tabular}{l l l l l l }
\hline\hline
                                              &      J2124-3358      &      J2129-5721      &      J2145-0750      &      J2150-0326      &      J2241-5236      \\
\hline
                  MJD range                   &    58441 - 59645     &    58441 - 59645     &    58441 - 59645     &    58659 - 59595     &    58441 - 59644     \\
       Right ascension (RA) [hh:mm:ss]        & \textbf{21:24:43.83534(4)} & \textbf{21:29:22.78125(3)} & \textbf{21:45:50.45355(5)} & \textbf{21:50:27.23531(10)} & \textbf{22:41:42.039596(4)} \\
        Declination (DEC) [dd:mm:ss]          & \textbf{-33:58:45.4778(9)} & \textbf{-57:21:14.3313(3)} & \textbf{-07:50:18.589(2)} & \textbf{-03:26:32.832(4)} & \textbf{-52:36:36.27108(4)} \\
     Proper motion in RA [mas $yr^{-1}$]      &  \textbf{-13.9(4)}   &   \textbf{9.3(2)}    &  \textbf{-10.3(8)}   &          --          &  \textbf{18.75(3)}   \\
    Proper motion in DEC [mas $yr^{-1}$]      &  \textbf{-49.8(8)}   &   \textbf{-9.7(3)}   &    \textbf{-7(2)}    &          --          &  \textbf{-5.17(5)}   \\
          Spin frequency [$s^{-1}$]           & \textbf{202.793896594889(3)} & \textbf{268.359230930218(3)} & \textbf{62.2958887629109(7)} & \textbf{284.84324670257(1)} & \textbf{457.3101563491688(8)} \\
    Spin frequency derivative [$s^{-2}$]      & \textbf{-8.463(2)$\cdot10^{-16}$} & \textbf{-1.5019(2)$\cdot10^{-15}$} & \textbf{-1.1563(5)$\cdot10^{-16}$} & \textbf{-6.622(9)$\cdot10^{-16}$} & \textbf{-1.44219(6)$\cdot10^{-15}$} \\
              DM [$cm^{-3}pc$]                &       4.592633       &      31.846624       &       9.002578       &      20.670000       &      11.410342       \\
               Parallax [mas]                 &    \textbf{3(1)}     &        0.4232        &   \textbf{3.4(7)}    &          --          &   \textbf{1.2(2)}    \\
                Binary model                  &          --          &         ELL1         &         ELL1         &         ELL1         &         ELL1         \\
             Orbital period [d]               &          --          & \textbf{6.625492997(2)} & \textbf{6.8389025093(9)} & \textbf{4.044550594(3)} & \textbf{1.4567223817(2)$\cdot10^{-1}$} \\
      Projected semi-major axis [lt-s]        &          --          & \textbf{3.5005689(4)} & \textbf{10.1641111(4)} & \textbf{3.320713(1)} & \textbf{0.02579534(6)} \\
          Epoch of periastron [MJD]           &          --          &          --          &          --          &          --          &          --          \\
        Longitude of periastron [deg]         &          --          &          --          &          --          &          --          &          --          \\
            Eccentricity of orbit             &          --          &          --          &          --          &          --          &          --          \\
        Epoch of ascending node [MJD]         &          --          & \textbf{59042.5329168(1)} & \textbf{59043.17563464(5)} & \textbf{59128.4793005(3)} & \textbf{59041.98680351(6)} \\
           First Laplace parameter            &          --          & \textbf{-0.0000043(2)} & \textbf{-0.00000691(8)} & \textbf{0.0000033(6)} & \textbf{-0.000004(5)} \\
          Second Laplace parameter            &          --          & \textbf{-0.0000118(2)} & \textbf{-0.00001808(7)} & \textbf{0.0000112(7)} & \textbf{-0.000003(5)} \\
      Time derivative of orbital period       &          --          &          --          &          --          &          --          &          --          \\
        Rate of advance of periastron         &          --          &          --          &      0.0242437       &          --          &          --          \\
 Rate of change of projected semi-major axis  &          --          &          --          &          --          &          --          &          --          \\
             Sine of inclination              &          --          &          --          &          --          &          --          &          --          \\
        Companion mass [$M_{\odot}$]          &          --          &          --          &          --          &          --          &          --          \\
    Shapiro delay parameter h3 [$\mu s$]      &          --          &          --          &          --          &          --          &          --          \\
    Shapiro delay parameter h4 [$\mu s$]      &          --          &          --          &          --          &          --          &          --          \\
\hline
\end{tabular}
\end{center}
\end{rotatetable*}
\end{table*}

\end{document}